\documentclass[11pt,a4paper]{article}
\pdfoutput=1

\usepackage{jheppub}

\usepackage{amsmath}
\usepackage{amssymb}
\usepackage{graphicx}
\usepackage{float}
\usepackage{braket}
\usepackage{hyperref}
\usepackage{multirow}
\usepackage{tikz}
\usepackage{pbox}
\usepackage{caption}
\usepackage{enumerate}
\usetikzlibrary{arrows,decorations.markings}

\usepackage[all]{hypcap}

\usepackage{feynmp}
\newcommand{\om}{\omega}

\newcommand{\dif}{\mathrm{d}}

\newcommand{\pt}{\partial}
\newcommand{\al}{\alpha}
\newcommand{\del}{\delta}

\newcommand{\abs}[1]{\lvert #1 \rvert}

\newcommand{\rr}{\ensuremath{\mathbb{R}}}
\newcommand{\cc}{\ensuremath{\mathbb{C}}}
\newcommand{\zz}{\ensuremath{\mathbb{Z}}}
\newcommand{\nn}{\ensuremath{\mathbb{N}}}
\newcommand{\affr}{\ensuremath{\mathrm{Aff} (\mathbb{R})^+}}

\usepackage{color}

\title{Quantum mechanics of a generalised rigid body}
\author{Ben Gripaios}
\author{and Dave Sutherland}

\affiliation{Cavendish Laboratory, \\ J.J.\ Thomson Avenue, Cambridge, UK}

\emailAdd{gripaios@hep.phy.cam.ac.uk}
\emailAdd{dws28@cam.ac.uk}

\preprint{Cavendish-HEP-15/03}

\abstract{We consider the quantum version of Arnold's generalisation of a
  rigid body in classical mechanics. Thus, we quantise the motion on
  an arbitrary Lie group manifold of a particle 
whose classical
  trajectories
correspond to the geodesics of any one-sided-invariant metric.
We show how the derivation of the spectrum of energy eigenstates can
be simplified by making use of automorphisms of the Lie algebra and
(for groups of Type I) by methods of harmonic analysis. We show how
the method can be extended to cosets, generalising the linear rigid rotor.
As examples, we consider all connected and simply-connected Lie groups up to dimension 3.
This includes the universal cover of the archetypical rigid body,
along with a number of new exactly-solvable models. We also discuss a possible application to the topical problem of
quantising a perfect fluid.}


\arxivnumber{1504.01406}

\begin{document} 
\maketitle
\section{Introduction \label{sec:int}}
Euler showed long ago that the classical mechanics of a free, rigid
body moving about a
fixed point ({\em e.g.\ }its centre of mass) simplifies, in that the
motion of the angular momentum relative to the body depends (via the
Euler equations) only on
its own initial value and not on the position of the body in
space. Moreover, the magnitude of this angular momentum is conserved. 

Rigid body motion can be thought of as the geodesic motion (with
respect to a one-sided-invariant metric) of a particle on the Lie group
manifold $SO(3)$. In 1966, Arnold \cite{Arnold} showed how Euler's simplifications generalise to
geodesic motion of a particle on {\em any} Lie group manifold $G$.
In a nutshell, the space of angular momenta of the rigid body is
replaced by the dual of the Lie algebra, the motion in that space
is governed by a set of generalised Euler-Arnold equations, and the motion is
restricted to an orbit of the coadjoint representation of $G$
(generalising the spheres of constant magnitude of angular momentum for
the rigid body). 

Arnold even went so far as to generalise to infinite-dimensional
Lie groups, such as the group, $\mathrm{Diff} (M)$, of diffeomorphisms
of a manifold $M$.\footnote{Such groups do not have the usual
  desiderata of finite-dimensional Lie groups: the exponential map does not necessarily exist for non-compact $M$, and even for
compact $M$ it may not be locally-onto. Indeed $\mathrm{Diff}
(\mathbb{R})$ and $\mathrm{Diff} (S^1)$, respectively, furnish counterexamples
\cite{Khesin}.} For the subgroup of volume-preserving diffeomorphisms,
the Arnold-Euler equations describe the motion of an incompressible
fluid on $M$; for the
particular case where $M$ is the two-torus, Arnold obtained a
celebrated result on the stability of fluid flows and thence an estimate of
the unreliability of long-term weather forecasting. 

The quantum mechanics of a rigid body has also been known for some time (see,
{\em e.g.}, \cite{casimir}). Here we attempt to carry out the generalisation to
arbitrary $G$, as well as to coset spaces $G/H$, where $H$ is a proper
subgroup of $G$. We then study, as illustrative examples, all connected and
simply-connected Lie groups up to dimension 3.\footnote{Groups
  that are connected but not simply-connected can be described within
  our formalism by choosing $H$ to be some discrete subgroup of the
  centre of $G$, the universal cover. If $H$ is also compact, the required harmonic
  analysis is straightforward. For groups that are not connected, one
  must first construct the irreps by induction from the component
  connected to the identity.}

It is too much to hope
that these quantum-mechanical systems are all exactly solvable,
because the archetypical case of an asymmetric rigid body already provides a
counterexample: there we can, at best, reduce the problem of finding the
energy spectrum to  diagonalization of finite-dimensional matrices.
But in many cases, we encounter systems that are exactly solvable,
using standard results in
representation theory and harmonic analysis.\footnote{In the
  unsolvable cases, we find that imposing
  an additional symmetry on the dynamics invariably results in an
  exactly-solvable model. Again, this is much like the rigid body,
  where the cylindrically- and spherically-symmetric cases are both
  solvable.} 
In this way, we uncover a number of new, exactly-solvable models in
quantum mechanics, generalising a rigid body.\footnote{Refs.\ \cite{PROP:PROP19790271102,1.533401} have
    studied the spectral properties of a restricted class of
    Hamiltonians (corresponding to bi-invariant metrics), using alternative path-integral and Green's
    function methods.} We hope that some of these
will turn out to be of physical interest. We also hope that our
results may provide some guidance towards the solution of a problem that is of some topical
interest, namely the quantisation of incompressible fluids. The quantisation of a
compressible fluid has been carried out using standard methods of
quantum field theory in
\cite{Endlich:2010hf,Gripaios:2014yha}. Classically, an incompressible
fluid is easily obtained as a limit of a compressible fluid, by
restricting to flows in which the fluid velocity is much lower than
the fluid's sound speed. One cannot impose such a restriction at the
quantum level.\footnote{To put it another way: because the sound wave modes are
  gapless, they cannot be integrated out to provide a low-energy,
  effective field theory of the incompressible modes.} Nevertheless,
a
quantum incompressible fluid may exist {\em sui generis}, and a
  generalisation to infinite-dimensional Lie groups of the approach we develop here may provide a means to describe it.

The outline is as follows. In the next Section, we discuss the
formalism. In \S\ref{sec:one} we quickly dispatch the 1-dimensional
Lie group $\rr$, where the dynamics is that of a free particle in
1-d and generalise to $\rr^n$. In \S\ref{sec:two}, we solve the
unique non-Abelian group in 2 dimensions, which is isomorphic to the group of
orientation-preserving affine transformations of $\rr$. There are two
proper subgroups isomorphic to $\rr$, up to conjugation, and we are
able to formulate and solve the quantum mechanics problem on the corresponding coset
space for one of these. In \S\ref{sec:three}, we study the groups that
arise in
dimension 3, ending with our first compact example, the universal
cover $SU(2)$ of the rigid body group, which is perhaps where we ought to have begun.
\section{Formalism \label{sec:for}}

Before we can attempt to solve the quantum mechanics of a
free particle on a Lie group $G$ or coset space $G/H$, we must first define it. A proper
definition of quantum mechanics is still lacking (and occupies much
learned discussion in the literature) and so we shall simply follow
our noses, in an attempt to come up with something suitable.
We begin by discussing the dynamics on a group $G$ and then generalise to
a coset space $G/H$.
\subsection{Quantum mechanics on $G$}
What are the necessary ingredients for the quantum mechanics of a free
particle moving on a Lie group manifold $G$? 

Firstly, we need a Hilbert space of physical states. Every Lie group
admits a unique (up to a multiplication by a constant) left (say) invariant
measure $\mu$ and so an obvious candidate for the Hilbert space for
the dynamics on $G$ is the
space $L^2(G,\mu)$ of (equivalence classes of) functions $f:G \rightarrow \cc$ that are
square-integrable w.r.t. $\mu$.
\footnote{In fact, from a physical point of view, we
also wish to include states that do not belong to
$L^2(G,\mu)$.
These generalised eigenvectors (also known as non-normalisable modes or
scattering states) can be thought of as living in the dual space to
the maximal subspace of $L^2(G,\mu)$ on which the
operators corresponding to observables are defined.
For $G = \rr$, for example, the generalised eigenfunctions of the
operators $x$ and $\pt_x$, correspond to
the Dirac delta function and $e^{ikx}$, respectively. 
The corresponding eigenvalues are included automatically in the functional-analytic
definition of the spectrum
as the complement in $\cc$ of the resolvent set.}

Secondly, we need some operators that act on the Hilbert space
$L^2(G,\mu)$ and that play the r\^{o}le of observables. 
The obvious candidates are the group elements themselves, which have
an action on $f \in L^2(G,\mu)$ by left multiplication of the argument of
the function. But these are not good candidates for observables when
the dynamics is that of a free particle! Indeed, consider the simplest
example of free-particle motion on the Lie group $\rr$, with
co-ordinate $x \in \rr$. The classical action is $\int dt
\frac{1}{2}\dot{x}^2$. If we attempt to compute the 2-point
correlation function of $x$ in Fourier space, we get $\langle x(t) x(t^\prime)
\rangle = \int \frac{\dif \om}{\om^2} e^{i\om t}$; because of the pole
at $\om = 0$, the Fourier transform is undefined. In the language of
physicists, the 2-point function is infra-red divergent, and so $x$
cannot be an observable.\footnote{A similar argument also shows that we
  cannot study the quantum mechanics on a Lie group $G$ by assuming
  that the energy eigenstates are localised near some $g \in G$ and
  then using the fact that $G$ is locally like $\rr^n$ to approximate
  the path integral as a Gaussian integral. As we shall see in
  explicit examples later on,
  the energy eigenstates correspond to wavefunctions that are
  completely delocalised in $G$.}

So, we need to find some other operators to play the r\^{o}le of
observables. To do so, we consider (via the exponential map), the
infinitesimal version of the group action described above. This gives
rise to right-invariant vector
fields sending some dense subspace (say, the smooth functions of compact support) of $L^2(G,\mu)$ into $L^2(G,\mu)$ and corresponding to formally self-adjoint
operators. Thus we
take a subset of the observables to be the space of right-invariant vector
fields on $G$, otherwise known as the Lie algebra, $\mathfrak{g}$, of $G$.\footnote{There is, of course, another reason for singling out
  $\mathfrak{g}$ as the space of observables: it is what we get in the
  archetypical example of a rigid body!} The quantum commutators of the
theory, which are defined by composition of the operators
corresponding to observables, are then just given by the Lie brackets.

We can enlarge the space of observables further by passing to
the universal enveloping algebra. This enables us to obtain our third
and final ingredient, which is a prescription for the dynamics, in the form of a
 Hamiltonian. Classically, this is easy to do. A Lie group is equipped with an
infinite family of metrics, namely the one-sided-invariant metrics. Picking one
such metric, a Hamiltonian (which is just the kinetic energy in this
free-particle case) is given by plugging the velocity vector
$\dot{g}$ corresponding to a classical trajectory $g(t) \in G$
into the metric.

Quantum mechanically, the corresponding Hamiltonian should be some operator in our
space of observables. Now the one-sided-invariant metrics on a group $G$ of
dimension $n$ are in
1-to-1 correspondence with the positive-definite, symmetric, $n \times
n$, real matrices. We thus take the Hamiltonian to be the quadratic
element of the universal enveloping algebra given by
\begin{gather}
\mathcal{H} = \sum_{i,j} \alpha_{ij} X_i^R X_j^R,
\end{gather}
where $\alpha_{ij}$ are the entries of such a matrix and where $\{
X^R_i\}$ is a basis for the Lie algebra of right-invariant vector fields.
This Hamiltonian is a Laplacian operator, which is formally self-adjoint 
with respect to the left-invariant Haar measure. Note that it is
  not, in general, equal to the Laplace-Beltrami operator constructed
  from the metric corresponding to $\alpha_{ij}$, nor is it a Casimir invariant for
  generic values of $\alpha_{ij}$. Nevertheless, it has three
  desirable properties that make it suitable as a quantum mechanical
  Hamiltonian. Firstly, it is a natural generalisation of what one obtains
  for the rigid body, where $\alpha_{ij}$ corresponds to the inverse
  of the moment of inertia tensor (see, \emph{e.g.}, \cite{GordyCook}). Secondly, it is explicitly written
  in terms of other physical observables, viz. the right-invariant
  vector fields. This makes it possible to build a comprehensive
  physical theory from it, which relates the results of measurements
  of distinct observables. Thirdly, being a symmetric, elliptic
  differential operator, it has a self-adjoint extension to
  $L^2(G,\mu)$. Even better, since we can perform a change of basis,
  $X_i^R \rightarrow X_i^{\prime R}$, such that $\mathcal{H}$ can be
  written as $\sum_i X^{\prime R}_i X^{\prime R}_i$, we have that
  $\mathcal{H}$ is both essentially self-adjoint (see, \emph{e.g.},~\cite{terelstrobinson}) and non-positive
  \cite{hoermander}. This guarantees that the spectrum is a (closed)
  subset of $[0,\infty)$, and so the energy is bounded below.\footnote{We remark that $\mathcal{H}$ may even enjoy the same properties when $\alpha_{ij}$ is degenerate, such that $\{X_i^{\prime R}\}$ do not span $\mathfrak{g}$, provided that they satisfy H\"{o}rmander's condition \cite{hoermander}, {\em viz.} that $\{X_i^{\prime R}\}$ and all Lie brackets formed recursively therefrom do span $\mathfrak{g}$.}

In all, $\mathcal{H}$ is a very plausible candidate for a
quantum-mechanical Hamiltonian. One could easily
generalise it further, by adding, for example, a term linear in
$X_i^R$. By analogy with the rigid body, this should presumably be
interpreted as corresponding to motion in a external field coupled to
the generalised angular momentum, generalising a magnetic field
coupling to angular momentum in the case of the rigid body.
\subsection{Quantum mechanics on $G/H$}
In the case of a (left, say) coset space, we can proceed in much the same way,
although there is an obstacle in that we may not be able to find a
$G$-invariant measure on $G/H$. Indeed, $G/H$ admits an
invariant measure iff. the restriction to $H$ of the modular function on $G$
coincides with the modular function on $H$ \cite{vilenkin}. \footnote{One way to
construct a measure is as follows
\cite{santalo}. Let $G$ and $H$ have dimension $n$ and $k$,
respectively and let $\{ \om^1, \dots, \om^{n-k}\}$ be a set of
linearly-independent left-invariant 1-forms on $G$ whose restriction
to $T_e(H)$ vanishes. The $(n-k)$-form $\om \equiv \om^1 \wedge \om^2 \dots
\wedge \om^{n-k}$ is unique up to multiplication by a constant and $G/H$
has a $G$-invariant volume form $\Omega$ (and hence a $G$-invariant
measure) iff. $d \om = 0$. If this condition holds, $\omega$ is given by pulling
back $\Omega$ using the natural projection $\pi: G \rightarrow G/H$.} In this paper we assume $H$ to be a closed Lie subgroup of $G$, and $\dim H < \dim G$. \footnote{In the case of finite $H$, its effect is to restrict the irreps appearing in the decomposition of the regular representation; consider $SU(2) / \mathbb{Z}_2 \cong SO(3)$ (\S\ref{sec:b8}) which only contains the irreps of integer spin.}

If we have a measure $\mu$, we can define
a representation of $G$ on functions in $L^2 (G/H,\mu)$ by the left
group action on the arguments of the functions, just as we did
before. From there, we can define formally self-adjoint observables
via the infinitesimal action of the exponential map. Equivalently, we
may simply take the right-invariant vector fields on $G$ and push them forward to $G/H$ using
the natural projection $\pi: G \rightarrow G/H$. This yields well-defined vector fields on $G/H$, since
right invariance of the vector fields on $G$ guarantees an identical
outcome when pushing forward the vectors at all points in the preimage
$\pi^{-1}(p) \subset G$ of a point $p\in G/H$.\footnote{Proof: Let $X$
  be a vector field on $G$ and let $X_g$ be its value at $g \in
  G$. These can be defined by their action on differentiable
  functions from $G$ to $\rr$. Now let $g^\prime \in G$ lie in the
  same coset as $g$, s.\ t.\ $g^\prime = gh$, for some $h \in H$. The pushforwards
  of $X_g$ and $X_{g^\prime}$ are given by $\pi^*(X_g) f =
  X_g f(\pi(g))$ (where $f: G/H \rightarrow \rr$ is differentiable) and $\pi^*(X_{g^\prime}) f =
  X_{g^\prime} f(\pi(g^\prime)) = X_{g^\prime} f(\pi(g)) = X_{gh}
  f(\pi(g))$. These are equal by right-invariance.}

As before, we then assume that the quantum Hamiltonian is just an
arbitrary symmetric, negative quadratic element in the universal enveloping algebra formed
from these vector fields and their Lie brackets.\footnote{Ref.~\cite{morozov1990} addresses the related question of identifying necessary conditions
for an element that is quadratic
in the generators of an arbitrary algebra to correspond to a
quantum-mechanical Hamiltonian; one case identified is that in which
the algebra corresponds to that of $G$-invariant vector fields on a coset space $G/H$ admitting a $G$-invariant metric, {\em viz.} the case discussed here.}
\subsection{Method of solution}
We now describe our strategy for solving these quantum mechanical
systems. The spectrum of energy eigenstates, for example, is
given by the spectrum of a Laplacian, so we could attempt to solve this partial
differential eigenvalue equation directly. We define the spectrum as the complement of the resolvent set of our Hamiltonian, and we can test for membership of this set by constructing Weyl sequences out of the solutions of the PDE. But as we shall see, these
PDEs are rather uninviting in their appearance and are not always amenable
to solution by humble methods such as separation of variables, nor are the Weyl sequences easy to construct.

One approach, explored in \cite{Avetisyan:2012zn}, would be to express
a generic function on $G$ in terms of the simultaneous
generalised eigenfunctions of one, or more, of the left-invariant vector
fields.\footnote{We thank Z. Avetisyan for discussions on this point.} The left-invariant vector fields commute everywhere with the
right-invariant vector fields, and by extension with the
Hamiltonian. Therefore, the substitution into the PDE of a simultaneous
eigenfunction of the left-invariant vector field(s) may reduce the
problem to an ODE. However, the generalised eigenfunctions may
themselves prove difficult to find. Even if they can be found, the
necessary Weyl sequences needed to find the spectrum remain difficult to construct. Even if the
spectrum can be found in this way, one is still lacking a crucial piece of
physical information, namely the spectral density. This is needed to
compute, {\em e.g.}, the thermal properties of such quantum-mechanical systems.

A more sure-fire approach is to try to exploit the group-theoretic and
geometric structure underlying the dynamics in a different way. Indeed,
the Hilbert space carries a highly reducible, unitary representation of $G$ (the
left regular representation) and so an obvious step is to try to
reduce this representation into its unitary, irreducible representations
(`unirreps', henceforth). The action of the Hamiltonian, which is an
element of the universal enveloping algebra, then
decomposes into actions on the irreducible subspaces, which we can attack individually.\footnote{This decomposition corresponds to the
  fact that the classical motion is confined to a coadjoint orbit
  of $G$. Each coadjoint orbit has an associated reduced phase
  space and mathematicians have used the presumed existence of a
  corresponding reduced quantum-mechanical system to find the unirreps
  of a variety of Lie groups. See \cite{kir04} for details.} Moreover, this decomposition constitutes a unitary map between the
  Hilbert spaces of the left regular rep and the unirreps. Thus, any
  Weyl sequence we may construct for an eigenfunction of the
  Hamiltonian in the latter space (which, being of lower dimension, is
  often easier) gives rise to a Weyl sequence for the solution
of the original PDE.

The decomposition of functions on $G/H$ generalises the Fourier transform on
$\rr$ and is a standard tool of harmonic analysis. Let us describe it
schematically here for the simplest case, where $H =
\{e\}$ and $G$ is unimodular. Let the unirreps of $G$ be $D^\lambda$, where $\lambda$
labels the equivalence classes.
Given a function $f$ in
$L^2(G,\mu_g)$ (where we put a subscript $g$ on $\mu$ to remind ourselves that
it is a measure on $G$), define an integral transform by
\begin{gather}\label{eq:ft}
\tilde{f} (\lambda) \equiv \int \mu_g f(g) \overline{D}^\lambda(g),
\end{gather}
where $\overline{D}$ denotes the Hermitian conjugate of $D$, and the operator $\tilde f(\lambda)$ acts on the Hilbert space of the irrep $\lambda$. There exists a measure
$\mu_\lambda$ on the space of equivalence classes of unirreps, such
that the transform may be inverted, to obtain
\begin{gather} \label{eq:inv}
f(g) = \int \mu_\lambda \mathrm{tr} \left[ D^\lambda(g) \tilde{f} (\lambda) \right].
\end{gather}
(As an example, for the case $G=\rr$ with $x \in \rr$, let the
invariant measure be $\mu_x = \dif x$. The irreps are 1 dimensional,
$D^\lambda (x) = e^{i\lambda x}, \lambda \in \rr$ and the operator $\tilde{f} (\lambda)$ is just the usual $\cc$-valued
Fourier transform. By the Parseval theorem, we
have that the Plancherel measure is $\mu_\lambda = \frac{\dif
  \lambda}{2\pi}$.)
The inversion is thus an explicit decomposition of $L^2(G,\mu_g)$ into
subspaces of operators acting on irreps $\lambda$, with measure
$\mu_\lambda$. Notice that the decomposition involves, in general, a
direct integral rather than a direct sum. For non-compact groups, some set of unirreps may have zero
Plancherel measure, meaning that they do not appear in the
decomposition. 

At this point, a triumvirate of problems appear. The first
is that, even when $G$ is non-compact (as we shall see explicitly in the dimension two example), we will
need to know all the unirreps of $G$ to decompose the left regular reps
on $G/H$ for arbitrary $H$.\footnote{For $H$ compact, we expect that $L^2(G/H)$ is isomorphic to the subspace of $L^2(G)$ defined by the functions that are constant within the cosets.
(For H non-compact, such functions are not in $L_2(G)$.) So all
the unirreps in $L^2 (G/H)$ already appear in $L^2(G)$.} 
The
classification of the unirreps of a general Lie group is an unsolved
problem\footnote{Examples of general cases where the classification is
  known are when $G$ is either a compact \cite{peterweyl} or an exponential Lie
  group \cite{currey}. An exponential Lie group is a special case of a solvable
  group, for which the exponential map is a diffeomorphism.} and we will only be able to make progress in this way in cases
where the irreps appearing in the decomposition are known. The second
is that the decomposition is not unique for certain Lie
groups, known as Type II groups; it is unclear how one should deal
with these. The third is that the unirreps of
non-Abelian groups are often multi-dimensional (or even infinite
dimensional in the non-compact case). Thus, even if we
manage to effect the decomposition, we may be left with a matrix or
differential operator representing the action of the Hamiltonian whose
eigenvalues we are unable to find. This is the obstruction that occurs
in the rigid body case and we shall see that it arises in several
other cases in dimension three.

Nevertheless, we shall see that many cases are tractable. Two
simplifying observations reduce the burden.
The first is that while a generic one-sided-invariant
Hamiltonian has $n(n+1)/2$ arbitrary coefficients,
 we may be able to simplify its form, w.\ l.\ o.\ g., using automorphisms of the Lie
algebra $\mathfrak{g}$. These are just the invertible linear maps from
$\mathfrak{g}$ to itself that preserve the structure constants of the
algebra (in
some basis) and we may use them to simplify the Hamiltonian without
changing the form of the operators representing the algebra elements.
Unfortunately, finding the automorphisms (which may be
outer) generally requires a brute-force approach.\footnote{Some
  simplifying tricks may be found in \cite{fisher}, which classifies the
  automorphism groups for all Lie algebras up to dimension five. This
  work also makes it clear that there is no simple way to treat direct
product groups, because one cannot use the automorphisms to
block-diagonalize the Hamiltonian into operators acting only on the
separate factors.}

The second observation is that there is a group, $K$,
of residual automorphisms
that preserve the form of the simplified Hamiltonian. If non-trivial, $K$ provides further constraints on the form of the
spectrum. Consider, for example, the maximal compact subgroup,
$\overline{K} \subset K$. We can simplify the problem of finding the
spectrum by decomposing the Hilbert space
into irreps of $\overline{K}$, because Schur's lemmata then imply that
the Hamiltonian $\mathcal{H}$ acts as the identity on the irreps of $\overline{K}$ and will not mix
inequivalent irreps appearing in the decomposition (it can mix
equivalent irreps, however).

In fact, it is sometimes convenient to consider an even larger subgroup
of automorphisms, namely those that preserve the form of the
simplified Hamiltonian up to a permutation of the latter's
coefficients. We call this the group of permuting automorphisms and
denote it by $J$. Evidently, there is a homomorphism from $J$ to the
permutation group of the coefficients with kernel $K$ and so $K$ is a
normal subgroup of $J$. To give an example of how
$J$ may be useful, given one energy eigenvalue,
corresponding to a state carrying an irrep of $J$, one can find other
energy eigenvalues of states in the irrep by effecting the corresponding permutations of the
coefficients. 
\section{Dimension one and $\rr^n$ \label{sec:one}}
In one dimension, the only Lie algebra is $[X,X]=0$, and
the corresponding (connected and simply-connected) group is isomorphic
to $\rr$. We might as well do $\rr^n, n \in \nn$, while we are at it. The
non-trivial automorphisms of the algebra are all outer and consist of all invertible linear maps from
$\rr^n$ to itself, {\em viz.} $GL(n,\rr)$; a transformation in
$SO(n,\rr) \subset GL(n,\rr)$ can be used to write the most general
Hamiltonian in diagonal form and then a diagonal matrix in $GL(n,\rr)$
can be used to write the Hamiltonian in the form
\begin{gather} \label{eq:Krr}
\mathcal{H} = X_1^2 + X_2^2 + \dots + X_n^2,
\end{gather}
where $\{X_1,\dots,X_n\}$ is some basis for the Lie algebra. 

The unirreps of the Abelian group $\rr^n$ are one-dimensional, with action on $\cc$
given by multiplication by $e^{i\vec{k} \cdot \vec{x}}$, with $\vec{k} \in \rr^n$.  The
decomposition of the left regular representation (in which the Lie
algebra acts as $X_i = \frac{\partial}{\partial x_i}$) into irreps is
achieved by means of the Fourier transform
\begin{gather}
f(\vec{x}) = \int \frac{\dif^n \vec{k}}{(2\pi)^n} e^{i\vec{k} \cdot
  \vec{x}} \tilde{f} (\vec{k}).
\end{gather}
The energy eigenvalues are thus given by $E = \vec{k} \cdot \vec{k}
\geq 0$, with Plancherel measure
$\frac{\dif^n \vec{k}}{(2\pi)^n}$.

In this case, then, we are easily able to find the spectrum without
recourse to the group, $O(n,\rr)$, of residual automorphisms of (\ref{eq:Krr}).
We could, of course, also decompose 
the left regular rep into irreps of the group $O(n,\rr)$, by passing
to hyperspherical polar co-ordinates, whereupon the decomposition
reduces to that of functions on the hypersphere.
This is hardly a simplification, since the only invariant subspace of 
both $\rr^n$ and $O(n,\rr)$ is spanned by the constant function.
There is one simplifying consequence, however, which is that the 
invariant subspaces of $\rr^n$ that are carried into one another by
the action of $O(n,\rr)$, namely all those with equal values of
$\vec{k} \cdot \vec{k}$, must be degenerate in energy, as indeed they are.
\section{Dimension two \label{sec:two}}
We now illustrate the decomposition of the left regular rep on $f \in
L^2(G,\mu)$ into irreps, and the subsequent derivation of the spectrum, in the case where $G$ is a Lie group of dimension
  2. There is a unique non-Abelian 2-d Lie algebra (up to isomorphism), with Lie
  bracket $[X,Y]=Y$ in some basis $\{X,Y\}$. The corresponding connected and
simply-connected group is
the group of orientation-preserving
affine transformations of the real line. We denote this by $G=\affr$, also
known as the `$ax+b$ group' with $a>0$. This group is non-unimodular
and so we must be careful to pick the left-invariant Haar measure for
$\mu$.

\subsection{$\affr$ group properties and irreps}

$\affr$ is isomorphic to the group of matrices
\begin{equation}
 \affr = \Big\{ \left( \begin{matrix} a & b \\ 0 & 1 \end{matrix} \right): a,b \in \rr, a > 0 \Big\} 
\label{eq:groupdef}
\end{equation}
under matrix multiplication. Its algebra is isomorphic to
\begin{equation*}
\Big\{ \left( \begin{matrix}
\alpha & \beta \\ 0 &0
\end{matrix} \right) : \alpha, \beta \in \rr \Big\} \equiv \{ \alpha  X+\beta
Y \},
\end{equation*}
where, as above, $\{X,Y\}$ form a basis with the canonical Lie bracket
$[X,Y]=Y$. The left- and right-invariant Haar measures are proportional to
$ \frac{\dif a \dif b}{a^2}$ and $ \frac{\dif a \dif b}{a}$,
respectively.

The right-invariant vector fields are
\begin{gather} \label{eq:invvec}
X^R = - a \pt_a - b\pt_b, \; Y^R = -\pt_b,
\end{gather}
and the left-invariant vector fields are
\begin{gather} \label{eq:invvecr}
X^L =  a \pt_a, \; Y^L = a\pt_b.
\end{gather}
Each of these yields the desired algebra $[X, Y] = Y$, with the Lie
brackets of
left-invariant fields with right-invariant fields vanishing.

The group is both non-Abelian and non-compact and has 2 infinite-dimensional unirreps
\begin{equation}
 D^\pm(s,t; \begin{smallmatrix} a \\ b \end{smallmatrix} ) = e^{\pm 2 \pi i b s} \del(a s - t) t,
\end{equation}
which act on functions $\phi(t) \in L^2(\rr^+,\frac{\dif t}{t})$ thus:
\begin{equation}
 \int_0^\infty \frac{\dif t}{t}  D^\pm(s,t; \begin{smallmatrix} a \\ b \end{smallmatrix} ) \phi(t) = e^{\pm 2 \pi i b s} \phi(a s) .
 \label{eq:irrepaction}
\end{equation}
There is also a family of 1 dimensional unirreps,
\begin{equation}
 D^c(\begin{smallmatrix} a \\ b \end{smallmatrix} ) = a^{2 \pi i c} ,
\end{equation}
labelled by $c \in \rr$, which act by multiplication on the Hilbert space $\cc$ \cite{kir94}.

\subsection{Decomposition of the left regular representation into irreps}
By analogy with the similarity transform that block diagonalizes a
finite dimensional reducible matrix representation, we seek a unitary
transformation from $L^2 (G,\mu)$ to functions on the unitary dual of $G$, which consists of equivalence
classes of $G$ irreps, and is equipped with the concomitant Plancherel measure. In picking an explicit basis for coordinates on the group manifold and its unitary dual, we will forgo many of the subtleties of harmonic analysis and instead derive our results in explicit examples by appealing to properties of the Euclidean Fourier transform. We caution the reader that similar arguments may not be applicable in the general case.

Following (\ref{eq:ft}), we first define integral
transforms\footnote{The integral transform in (\ref{eq:ft}) is an operator on the Hilbert space of irreps; here we transform the `operator's matrix elements', with `indices' $s,t$ and measures $ds/s,dt/t$ instead.}\footnote{The extra factor
  of $\sqrt{t}$ arises because of a subtlety
  in the harmonic analysis of this and other non-unimodular
  groups. For more details, see
  \cite{Khalil,duflomoore}.}
\begin{align*}
\tilde f^p(s,t) &=  \int \frac{\dif a \dif b}{a^2} \overline{D}^p(s,t;\begin{smallmatrix} a \\ b \end{smallmatrix}) \sqrt{s} f(\begin{smallmatrix} a \\ b \end{smallmatrix}) ,\\
\tilde f^c &=  \int \frac{\dif a \dif b}{a^2} \overline{D}^c(\begin{smallmatrix} a \\ b \end{smallmatrix}) f(\begin{smallmatrix} a \\ b \end{smallmatrix}) ,
\end{align*}
where $p\in \{\pm\}$ and $\overline{D}(s,t) \equiv D^*(t,s)$ denotes the Hermitian conjugate of
$D$. Note that the desired group transformations of the $\tilde f$s
are fixed by the properties of the representation matrices $D(g)$ and
the Haar measure $\mu_g$. Schematically, the action $T(g^\prime)$ of the group element $g^\prime$ is:
\begin{align*}
 T(g^\prime) \tilde f^p(s,t) &=  \int \mu_g \overline{D}^p(s,t;g) \sqrt{s} f((g^\prime)^{-1} g)  \\
 &= \int \frac{\dif u}{u} \overline{D}^p(u,t;g^\prime) \int \mu_{(g^\prime)^{-1} g}  \overline{D}^p(s,u;(g^\prime)^{-1} g) \sqrt{s} f((g^\prime)^{-1} g) \\
 &= \int \frac{\dif u}{u} \tilde f^p(s,u) \overline{D}^p(u,t;g^\prime)  ,
\end{align*}
and thus the $\tilde f^p$ (and similarly the $\tilde f^c$) indeed transform under an irrep of $G$.

It is then the inversion of the above integral transforms --- the reconstruction of the original $f(\begin{smallmatrix} a \\ b \end{smallmatrix})$ from the $\tilde f^p(s,t)$ and $\tilde f^c$ --- that determines the weights with which each irrep appears in the left regular representation. To this end we observe that, using the counting measure $\sum_{p=\pm}$ over the two infinite dimensional irreps, it is possible to obtain a resolution of the identity on the group coordinates:\footnote{The equivalent relation on the irreps is just the Schur orthogonality relation
\begin{equation*}
 \int \frac{\dif a \dif b}{a^2} \overline{D}^p(t,s;\begin{smallmatrix} a \\ b \end{smallmatrix}) \sqrt{t} D^{p^\prime}(s^\prime,t^\prime;\begin{smallmatrix} a \\ b \end{smallmatrix}) \sqrt{t^\prime}
= \delta^{p p^\prime} \delta(s-s^\prime) \delta(t-t^\prime) s t.
\end{equation*}
}
\begin{equation*}
 \sum_{p=\pm} \int \frac{\dif s \dif t}{s t} D^{p}(s,t;\begin{smallmatrix} a^\prime \\ b^\prime \end{smallmatrix}) \sqrt{t} \overline{D}^p(t,s;\begin{smallmatrix} a \\ b \end{smallmatrix}) \sqrt{t}
= \delta(a-a^\prime) \delta(b-b^\prime) a^2 ,
\end{equation*}
and we can therefore write the inversion as
\begin{equation}
 \label{eq:decomp}
 f(\begin{smallmatrix} a \\ b \end{smallmatrix}) = \sum_{p=\pm} \int \frac{\dif s \dif t}{s t} D^{p}(s,t;\begin{smallmatrix} a \\ b \end{smallmatrix}) \sqrt{t} \tilde f^p(t,s) .
\end{equation}
$\sum_{p=\pm}$ is the Plancherel measure on the irreps, and is notably zero on the 1-d $c$ irreps, which do not feature in the decomposition of the left regular representation. Again, we explicitly compute the group action on our formula. Under left translation by $(g^\prime)^{-1}$
\begin{align*}
f((g^\prime)^{-1} g) &= \sum_{p=\pm} \int \frac{\dif s \dif t}{s t} D^{p}(s,t;(g^\prime)^{-1} g) \sqrt{t} \tilde f^p(t,s) \\
&=\sum_{p=\pm} \int \frac{\dif s \dif t}{s t} \frac{\dif u}{u} D^p(s,u;(g^\prime)^{-1} ) D^p(u,t; g ) \sqrt{t} \tilde f^p(t,s) \\
&= \sum_{p=\pm} \int \frac{\dif u \dif t}{u t} D^p(u,t; g ) \sqrt{t} \Big( \int \frac{\dif s}{s} \tilde f^p(t,s) D^p(s,u; g^\prime )  \Big) ,
\end{align*}
it is clear that the $\tilde f^p(s,t)$ which comprise $f(\begin{smallmatrix} a \\ b \end{smallmatrix})$ transform under the irrep action (\ref{eq:irrepaction}).

\subsection{The spectrum of the left regular representation\label{sec:leftspec}}
Following the rubric of \S \ref{sec:for}, the quantum Hamiltonian is
given by $\mathcal{H} = a_{ij} X_i^R X_j^R $, where $X^R_1 \equiv X^R$
and $X_2^R \equiv Y^R$ are the right-invariant vector fields
defined in (\ref{eq:invvec}). To reduce clutter, we drop the $R$ superscripts in what follows. 

To find the automorphisms of the algebra, consider an arbitrary,
invertible, linear transformation $M \in GL(\dim G,\rr)$ of the Lie algebra $\mathfrak{g}$, and require that $[M \xi, M \eta] = M[\xi,\eta], \, \forall \xi,\eta \in \mathfrak{g}$, {\em i.e.}, it leaves the structure
constants (namely the relation $[X,Y]=Y$, and the other trivial Lie
brackets) invariant. Thus, let $M_{ij} = \left( \begin{matrix} a
    & b \\ c & d \end{matrix} \right)$. From the invariance of
$[X_i,X_j]=f^k_{ij} X_k$, we obtain the set of constraints $M_{ip}
M_{jq} f^k_{ij} (M^{-1})_{rk} = f^r_{pq}$, where
$f^2_{12}=-f^2_{21}=1$ and all others are zero, and we sum over repeated indices. This yields two
non-trivial equations, $(ad-bc)b=0$ (implying $b = 0$, since $M$ is non-singular), and $a=1$, leaving
\begin{equation}
 M = \left( \begin{matrix} 1 & 0 \\ c & d \end{matrix} \right), d \neq 0 .
\end{equation}
Under conjugation by the matrix
$(\begin{smallmatrix} 0 & 1 \\ 1 & 0 \end{smallmatrix})$, these take
the same form as (\ref{eq:groupdef}) without the restriction that $d$
be positive: the group of automorphisms is therefore isomorphic to
$\mathrm{Aff}(\rr)$. For general $\mathcal{H} = \alpha X^2 + \beta \{X,Y\} + \gamma Y^2$, there exists a transformation $M=\left( \begin{smallmatrix} 1 &
    0 \\ \frac{\beta}{\sqrt{\alpha \gamma - \beta^2}} &
    \frac{\gamma}{\sqrt{\alpha \gamma - \beta^2}} \end{smallmatrix} \right)$ to put the
Hamiltonian in the simplified form $\mathcal{H} = A(X^2 + Y^2)$, $A>0$, which we will assume hereafter.

We are now in a position to solve for the spectrum of the left regular
representation, which is the set of eigenvalues $-E$ of the
Hamiltonian\footnote{We write $-E$ as we have used the mathematicians'
  convention of skew-Hermitian group generators, leading to a negative
  definite spectrum. We expect $E \geq 0$.}
\begin{equation*}
 \mathcal{H} f = A ( (a \pt_a + b \pt_b)^2 + ( \pt_b)^2) f = -E f .
\end{equation*}
This PDE looks hard to solve (in particular, it cannot be solved by the method of separation of variables using the coordinates $(a,b)$), but its solution is made simple by using the group structure of
(\ref{eq:decomp}) and integrating by parts:\footnote{One way of doing this would be to integrate over the delta function in $D^{p}(s,t;\begin{smallmatrix} a \\ b \end{smallmatrix})$ in the decomposition (\ref{eq:decomp})
\begin{equation*}
 f(\begin{smallmatrix} a \\ b \end{smallmatrix}) = \sum_{p=\pm} \int \frac{\dif t}{\sqrt{t}} e^{2 \pi i p \frac{b t}{a}} \tilde f^p(t,\frac{t}{a}) ,
\end{equation*}
whence, for example,
\begin{align*}
 a \pt_a f(\begin{smallmatrix} a \\ b \end{smallmatrix}) = \sum_{p=\pm} \int \frac{\dif t}{\sqrt{t}} e^{2 \pi i p \frac{b t}{a}} [ -2 \pi i p \frac{b t}{a} \tilde f^p(t,\frac{t}{a}) -\frac{t}{a} \tilde f^{p (0,1)}(t,\frac{t}{a}) ] \\
 = \sum_{p=\pm} \int \frac{\dif s \dif t}{s t} D^{p}(s,t;\begin{smallmatrix} a \\ b \end{smallmatrix}) \sqrt{t} [- 2 \pi i p b s - s \frac{\dif}{\dif s} ] \tilde f^p(t,s) ,
\end{align*}
$\tilde f^{p (0,1)}$ denoting a derivative with respect to the second argument.
}
\begin{equation*}
 \sum_{p=\pm} \int \frac{\dif s \dif t}{s t} D^{p}(s,t;\begin{smallmatrix} a \\ b \end{smallmatrix}) \sqrt{t} \left[ A( s^2 \frac{\dif^2}{\dif s^2} + s \frac{\dif}{\dif s} - 4 \pi^2 s^2 ) + E \right]\tilde f^p(t,s) = 0.
\end{equation*}
The resulting ODE is a modified Bessel equation, whose solutions are
well-known \cite{DLMF}. We select from these only the generalised eigenfunctions which are the limit of the Weyl sequences constructed in Appendix~\ref{app:weylseq} (which in turn yields generalised eigenfunctions of $L^2(\affr, \frac{\dif a \dif b}{a^2})$ after recomposition). The eigenvalues and eigenfunctions are therefore
\begin{equation}
 E \in [0,\infty); \quad \tilde f^p(t,s) = \tilde K_{\sqrt{\frac{E}{A}}} (2 \pi s) g(t), \, g(t) \in L^2(\rr^+,\frac{\dif t}{t}) \text{ arbitrary.}
\end{equation}
$\tilde K_\nu(x)$ is a modified Bessel function of imaginary order
$\nu$, behaving like $\sin (\nu \log x)$ near the origin, but exponentially decaying as $x
\to +\infty$ \cite[10.45]{DLMF}. For $g(t)$, we can choose a basis of
generalised functions, such as $\sqrt{t} e^{2 \pi i \lambda t}, \lambda \in \rr$. Thus
$\lambda$ (together with $p \in \{\pm\}$) parametrises a set of
degenerate generalised eigenfunctions.

Out of curiosity, and as a final check, we can recompose our solutions to
the Laplacian in terms of the group coordinates $a$ and $b$. Let, say, $\tilde f^+(t,s) = \tilde
K_\nu (2 \pi s) \sqrt{t} e^{2 \pi i \lambda t}$ and $\tilde f^-(t,s) = 0$, where $\nu = \sqrt{\frac{E}{A}}$ and $\lambda \in \rr$. Then,
\begin{align*}
 f(\begin{smallmatrix} a \\ b \end{smallmatrix}) &= \int \frac{\dif s \dif t}{s t} D^{+}(s,t;\begin{smallmatrix} a \\ b \end{smallmatrix}) \sqrt{t} \tilde K_\nu (2 \pi s) \sqrt{t} e^{2 \pi i \lambda t} ,\\
 &= a \int \dif s \, e^{2 \pi i (b+\lambda a) s} \tilde K_\nu (2 \pi s) ,\\
 & \propto a  \left(1 - i(b+\lambda a) \right)^{-1+i \nu} \frac{1}{F} \left( \left(1-F\right)^{2 i \nu} - \left(1+F\right)^{2 i \nu} \right) \text{ where } F(\lambda) = \sqrt{\frac{i(b+\lambda a)+1}{i(b+\lambda a)-1}}.
\end{align*}
We have explicitly checked that
this is a generalised eigenfunction of the Laplacian,  with eigenvalue $-A \nu^2 =
-E$. 

\subsection{The spectrum of the coset representation}
\begin{figure}
\begin{center}
\begin{tikzpicture}[scale=1.5]
\draw[thick,->] (-3,0) -- (3,0) node[anchor=west] {$b$};
\draw[thick,->] (0,0) -- (0,2) node[anchor=south] {$a$};
\draw[dashed] (-3,0.7) -- (3,0.7);
\draw[dashed] (-3,1.7) -- (3,1.7);
\fill (0,0.7) circle[radius=0.07] {};
\fill (0,1.7) circle[radius=0.07] {};
\fill[gray] (-1.5,1.7) circle[radius=0.07] {};
\draw[draw=black,thick,-triangle 90]  (-0.1,0.8) -- (-1.4,1.6) node[pos=0.5,below left]{$g$};
\draw[draw=black,thick,-triangle 90]  (-1.4,1.8) -- (-0.1,1.8) node[pos=0.5,above]{$h^{-1}$};
\end{tikzpicture}
\end{center}
\caption{The coset structure on $G/H$ for $\affr$. All points of equal $a$ belong to the same coset, indicated by the dashed lines. We choose points along the $a$ axis (such as the black circles) to form our coset representative. The group action on the coset representative is left multiplication by $g \in G$, followed by the projection of the result onto its coset representative by right multiplication by $h^{-1}$.\label{fig:affrcoset}}
\end{figure}
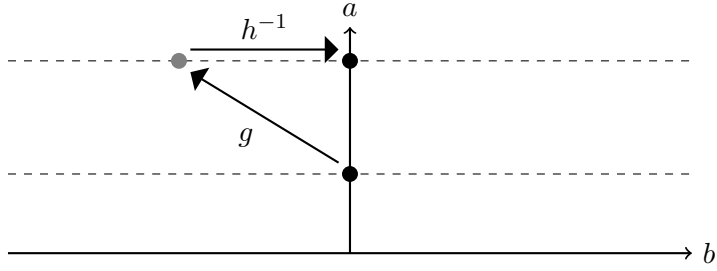

We now consider quantum mechanics on the quotients of $\affr$ by subgroups isomorphic to $\rr$. There are two such subgroups up to
conjugation, namely $H = \{ \left( \begin{matrix} 1 & b \\ 0 &
    1\end{matrix} \right)\}$, for which suitable coset representatives are $G/H = \{ \left( \begin{matrix} a & 0 \\ 0 &
    1\end{matrix} \right)\}$, illustrated in Fig.~\ref{fig:affrcoset}, and 
$H^\prime = \{ \left( \begin{matrix} a & 0 \\ 0 &
    1\end{matrix} \right)\}$, for which suitable coset representatives are $G/H = \{ \left( \begin{matrix} 1 & b \\ 0 &
    1\end{matrix} \right)\}$.\footnote{Note that, as expected from the
    arguments of \S\ref{sec:for}, the
    pushforwards of right-invariant fields, given by $\pi^*(X^R) = -a
    \pt_a, \pi^*(Y^R) = 0$ and $\pi^{\prime *}(X^R) = -b \pt_b,
    \pi^{\prime *}(Y^R) = - \pt_b$, are well-defined on $G/H$ and
    $G/H^\prime$, respectively, but that the pushforward $\pi^{\prime
      *}(Y^L) =  a\pt_b$ is not.} Then the group actions on $G/H$ and $G/H^\prime$, obtained by
multiplying on the left by $\left( \begin{matrix} g_a & g_b \\ 0 &
  1 \end{matrix} \right) \in G$, are $a \mapsto g_a a$ for $G/H$ and $b
\mapsto g_a b + g_b$ for $G/H^\prime$.

To formulate quantum mechanics we require a $G$-invariant measure.
The modular function on $G$ is $\Delta (a,b) = 1/a$. Both $H$ and
$H^\prime$ are unimodular, so only $H$ (with $a=1$)  yields an
invariant measure on $G/H$, given by $\frac{\dif a}{a}$.

Imbued with a group action, the functions $f(\Sigma) \in L^2(G/H,\dif
\mu)$ form a representation of $G$, which may similarly be decomposed
into irreps, following the
  same logic as before. We
  first define an integral transform of functions in this space
\begin{equation*}
 \tilde f^c =  \int \frac{\dif a}{a} \overline{D}^c(\begin{smallmatrix} a \\ 0 \end{smallmatrix}) f(a).
\end{equation*}
Now the appropriate decomposition of the identity is
\begin{equation*}
 \int \dif c \; D^{c}(\begin{smallmatrix} a^\prime \\ 0 \end{smallmatrix}) \overline{D}^c(\begin{smallmatrix} a \\ 0 \end{smallmatrix}) 
= a \delta(a-a^\prime) ,
\end{equation*}
implying the inversion theorem
\begin{equation*}
 f(a) = \int \dif c D^{c}(\begin{smallmatrix} a \\ 0 \end{smallmatrix}) \tilde f^c.
\end{equation*}
From this we read off that the coset representation on $L^2(G/H,\frac{\dif a}{a})$ decomposes into a direct
integral of the 1-d
irreps labelled by $c$, with Plancherel measure $\dif c$. 

Note that the irreps appearing in the coset rep on $G/H$ are
distinct from those appearing in the left regular rep on $G$ (where
the Plancherel measure is given by the counting measure on the irreps $\pm$). As we
remarked earlier, this can only happen because $H$ is non-compact.

The most general Hamiltonian is $\mathcal{H} = A [\pi^*(X^R)]^2$ and the eigenvalue equation reduces to:
\begin{equation*}
 [ A (a^2 \pt_a^2 + a \pt_a) + E ] f(a) = \int \dif c D^{c}(\begin{smallmatrix} a \\ 0 \end{smallmatrix}) [ - A (2 \pi c)^2 + E ] \tilde f^c = 0,
\end{equation*}
with a spectrum $E= A(2 \pi c)^2 \in [0,\infty)$ and uniform measure
on $c$.

\subsection{Solution of the PDEs via simultaneous eigenfunctions}

It is useful to compare these results with what is obtained via
solution of the PDEs via simultaneous (generalised) eigenfunctions. Indeed, the PDEs above can be reduced to second-order ODEs by
substituting a functional form that is simultaneously an eigenfunction
of a vector field that commutes with the Hamiltonian. This is most
simply illustrated in the case of the right regular representation
(with the right Haar measure $\frac{\dif a \dif b}{a}$), where the
Hamiltonian $\mathcal{H}$ is formed from the left-invariant fields $X^L =  a \pt_a$ and $Y^L = a\pt_b$:
\begin{gather}
\mathcal{H} = A\left[ (a\pt_a)^2 + (a\pt_b)^2 \right] .
\end{gather}
The Hamiltonian commutes by construction with $Y^R = \pt_b$, whose eigenfunctions are proportional to $e^{i \lambda b}$, $\lambda \in \rr$. Substitution of $f(\begin{smallmatrix} a \\ b \end{smallmatrix}) = e^{i \lambda b} g(a)$ into $\mathcal{H} f = - E f$ reduces the problem to an ODE
\begin{equation}
a^2 g^{\prime \prime} + a g^\prime + (\frac{E}{A} - a^2 \lambda^2) g = 0,
\end{equation}
with solutions for $E \geq 0$ of
\begin{gather}
 e^{ i \lambda b} \tilde{K}_{\sqrt{E/A}} (|\lambda|a).
\end{gather}
The spectrum is identical to that of the unitarily equivalent left regular
representation, and so these generalised eigenfunctions are
demonstrably complete by comparison with the results obtained via the
Fourier transform above. The left regular representation's own PDE
also simplifies to an ODE upon substitution of $f(\begin{smallmatrix}
  a \\ b \end{smallmatrix}) = a^q g(b)$, which is a simultaneous
generalised eigenfunction of $X^L = a \pt_a$. 
The generalised eigenfunctions that are obtained by solving the
resulting ODE are not the same as those obtained previously in \S
\ref{sec:leftspec}; this should come as no surprise, given that the
generalized eigenfunctions corresponding to a given eigenvalue are
infinitely degenerate.

\section{Dimension three \label{sec:three}}

Up to isomorphism, there are nine classes, I--IX, of 3-d Lie algebras (two of which contain
infinitely many algebras), which
were first catalogued by Bianchi \cite{bianchi,bianchi2}. Bianchi I is
isomorphic to $\mathbb{R}^3$, {\em cf. } \S \ref{sec:one}. Bianchi
II--VII all take the form of a semi-direct product group $\rr^2
\rtimes \rr$ and can be treated in a unified way; we treat the Heisenberg group (Bianchi II) in detail
below, in analogue to our treatment of $\affr$ in \S
\ref{sec:two}. Within the treatment of the Heisenberg case, references
are made to Tables~\ref{tab:groupstructure} and \ref{tab:spectrum}, from which
the reader may substitute in the corresponding results for his or her
favourite Bianchi algebra II--VII. 

Many of these results come from \cite{Avetisyan:2012zn} (though our
conventions differ by minus signs), in
which the unitary duals are obtained by induction via the Mackey
machine and the Plancherel measures are derived. Ref.~\cite{Avetisyan:2012zn}
also contains a result on the spectrum of a symmetric quadratic form of (left- or right-)invariant vectors, which would correspond
to the spectrum of our
quantum-mechanical Hamiltonian. Unfortunately, the result given is
incorrect. The erroneous result is that
(in our sign convention) the spectrum is given by $[-c, \infty)$, where $c<0$ is the constant
scalar Riemann curvature of the metric. The result is obtained using a result of Donnelly
\cite{donnelly}, also shown in \cite{mckean1970}, that applies only to the Laplace-Beltrami operator on manifolds whose sectional
curvatures are all non-positive. Unfortunately, neither is the operator in question the Laplace-Beltrami operator, nor are the sectional curvatures of the Bianchi groups all non-positive. Indeed, it is easy to furnish a counterexample: the Bianchi II group
has Haar measure $\dif x \dif y \dif z$ in our co-ordinates and the Hamiltonian is purely derivative. Thus, the constant
function on the group is a generalised eigenfunction, and the spectrum
includes $0 < -c$. Many further counterexamples appear below.

$SL(2,\rr)$ (Bianchi VIII) and $SU(2)$ (Bianchi IX) are discussed
separately, in \S\S \ref{sec:b9}--\ref{sec:b8}.

\subsection{Bianchi II--VII}

Denote by $H$ the Heisenberg group formed by the set of unit determinant upper triangular $3 \times 3$ real matrices
\begin{equation*}
 H = \Bigg\{ \left( \begin{matrix} 1 & z & y \\ 0 & 1 & x \\ 0 & 0 & 1 \end{matrix} \right): x,y,z \in \rr \Bigg\} 
\end{equation*}
under matrix multiplication. Note that the group composition law is a semi-direct product of those of $(x,y) \in \rr^2$ and $z \in \rr$: $(x,y,z) \circ (x^\prime,y^\prime,z^\prime) = (x+x^\prime,y+y^\prime+z x^\prime,z+z^\prime)$ (Table~\ref{tab:groupstructure}a). The corresponding algebra is then:
\begin{equation*}
\Big\{ \left( \begin{matrix} 0 & \gamma & \beta \\ 0 & 0 & \alpha \\ 0 & 0 & 0 \end{matrix} \right) , \alpha,\beta,\gamma \in \rr \Big\} \equiv \alpha  X + \beta Y + \gamma Z,
\end{equation*}
with the Lie bracket $[Z,X]=Y$. The group is unimodular and has Haar measure $ \dif x \dif y \dif z$ (the other Bianchi groups are not unimodular and in general we use the left-invariant Haar measure $\det F(-z) \dif x \dif y \dif z$, see Table~\ref{tab:groupstructure}a).

$H$ has a continuum of inequivalent infinite-dimensional irreps
labelled by $k \in \rr / \{0\}$
\begin{equation}
 D^k(s,t; \begin{smallmatrix} x \\ y \\ z \end{smallmatrix} ) = e^{i k (x s + y)} \del(s + z - t),
\end{equation}
which act on functions $\phi(t) \in L^2(\rr,\dif t)$ (Table~\ref{tab:groupstructure}b). There are also one dimensional irreps, labelled by $(\mu,\nu) \in \rr^2$,
\begin{equation}
 D^{\mu,\nu}(\begin{smallmatrix} x \\ y \\ z \end{smallmatrix} ) = e^{ i (\mu z + \nu x)} ,
\end{equation}
which act by multiplication on the Hilbert space $\cc$ \cite{kir94}.

As before, to decompose $L^2(H)$ we require completeness relations on
both the irrep (or unitary dual) coordinates, and the group coordinates. The former comes from the Schur orthogonality relation:
\begin{equation*}
 \int \dif x \dif y \dif z \overline{D}^k(t,s;\begin{smallmatrix} x \\ y \\ z \end{smallmatrix}) D^{k^\prime}(s^\prime,t^\prime;\begin{smallmatrix} x \\ y \\ z \end{smallmatrix})
= \delta(k - k^\prime) \delta(s-s^\prime) \delta(t-t^\prime) \frac{4 \pi^2}{\abs{k}} .
\end{equation*}
The $\frac{4 \pi^2}{\abs{k}}$ suggests that in the latter we should use the Plancherel measure $\frac{\abs{k}}{4 \pi^2}$ (Table~\ref{tab:groupstructure}c)
\begin{equation*}
 \int \dif k \dif s \dif t \frac{\abs{k}}{4 \pi^2} D^{k}(s,t;\begin{smallmatrix} x^\prime \\ y^\prime \\ z^\prime \end{smallmatrix}) \overline{D}^k(t,s;\begin{smallmatrix} x \\ y \\ z \end{smallmatrix}) 
= \delta(x-x^\prime) \delta(y-y^\prime) \delta(z-z^\prime) ,
\end{equation*}
to achieve the correct resolution of the identity w.r.t the Haar measure $\dif x \dif y \dif z$. Note again that the 1d $(\mu,\nu)$ irreps do not feature in the decomposition of $L^2(H)$, which takes the form:\footnote{Generally,
\begin{align*}
\tilde f^{k(,k_2)}(s,t) &=  \int \dif x \dif y \dif z \det F(-z) \overline{D}^{k(,k_2)}(s,t;\begin{smallmatrix} x \\ y \\ z \end{smallmatrix}) \det F(\frac{1}{2} s) f(\begin{smallmatrix} x \\ y \\ z \end{smallmatrix}) , \\
 f(\begin{smallmatrix} x \\ y \\ z \end{smallmatrix}) &= \sum_{k_2} \int \dif k \dif s \dif t \frac{\nu(k,k_2)}{4 \pi^2} D^{k(,k_2)}(s,t;\begin{smallmatrix} x \\ y \\ z \end{smallmatrix}) \det F(\frac{1}{2} t) \tilde f^{k(,k_2)}(t,s) ,
\end{align*}
$\nu(k,k_2)$ being the Plancherel measure of Table~\ref{tab:groupstructure}c.}
\begin{align*}
\tilde f^k(s,t) &=  \int \dif x \dif y \dif z \overline{D}^k(s,t;\begin{smallmatrix} x \\ y \\ z \end{smallmatrix}) f(\begin{smallmatrix} x \\ y \\ z \end{smallmatrix}) , \\
 f(\begin{smallmatrix} x \\ y \\ z \end{smallmatrix}) &= \int \dif k \dif s \dif t \frac{\abs{k}}{4 \pi^2} D^k(s,t;\begin{smallmatrix} x \\ y \\ z \end{smallmatrix}) \tilde f^k(t,s) .
\end{align*}

The Hamiltonian in the left regular rep is a quadratic form $a^{ij} X_i X_j$ of the Lie algebra components $X_j = (X,Y,Z)^T$, which may be simplified by the automorphism $M_{ij}$, where
\begin{equation*}
 M_{ij} = \left( \begin{matrix} a & 0 & c \\ d & aj-cg & f \\ g & 0 & j \end{matrix} \right), \quad a,c,d,f,g,j \in \rr, \quad aj-cg \neq 0 .
\end{equation*}
Without loss of generality, we may then write $\mathcal{H} = X^2 + A
Y^2 + Z^2$ (Table~\ref{tab:groupstructure}d,f). All automorphisms
$M_{ij}$ of the algebra induce an action (also an automorphism) on the
group coordinates $e^{\alpha_i X_i} \to e^{\alpha_i M_{ji} X_j}$, and,
through the $D$ matrices, also an action on the irrep coordinates,
whereby unirreps are mixed. Unlike \affr, there is a
non-trivial group of residual automorphisms, locally isomorphic to $SO(2,\rr)$, that preserve the simplified form of the Heisenberg Hamiltonian, viz. (Table~\ref{tab:groupstructure}e),
\begin{equation*}
 \lbrace \left( \begin{matrix} \cos \theta & 0 & \sin \theta \\ 0 & 1 & 0 \\ -\sin \theta & 0 & \cos \theta \end{matrix} \right),\left( \begin{matrix} 1 & 0 & 0 \\ 0 & -1 & 0 \\ 0 & 0 & -1 \end{matrix} \right) \rbrace, \theta \in [0,2 \pi).
\end{equation*}
These only mix unirreps with identical spectra, identifying a possible degeneracy. These are the pairs of irreps labelled by $k$ and $-k$, as we will see presently.

The actions of the Lie algebra elements are $X^R f(\begin{smallmatrix} x \\ y \\ z \end{smallmatrix}) = \pt_x f(\begin{smallmatrix} x \\ y \\ z \end{smallmatrix})$, $Y^R f(\begin{smallmatrix} x \\ y \\ z \end{smallmatrix}) = \pt_y f(\begin{smallmatrix} x \\ y \\ z \end{smallmatrix})$ and $Z^R f(\begin{smallmatrix} x \\ y \\ z \end{smallmatrix}) = (\pt_z + x \pt_y) f(\begin{smallmatrix} x \\ y \\ z \end{smallmatrix})$. This yields an eigenvalue equation:
\begin{align*}
 [\mathcal{H}+E] f &= [ \pt_x^2 + A \pt_y^2 + (\pt_z+x \pt_y)^2 + E ] f \\
 &= \int \dif k \dif s \dif t \frac{\abs{k}}{4 \pi^2} D^k(s,t;\begin{smallmatrix} x \\ y \\ z \end{smallmatrix}) \left[ \frac{\dif^2}{\dif s^2} + E - k^2 (A+s^2) \right]\tilde f^k(t,s) \\
 &= 0.
\end{align*}
The ODE (Table~\ref{tab:spectrum}a), which is just the Schr\"odinger equation for the simple harmonic oscillator, is solved by parabolic cylinder functions \cite{DLMF}, of which the normalisable ones reduce to (Table~\ref{tab:spectrum}b,c):
\begin{align*}
 E = (2 n+1) \abs{k} + k^2 A, n \in \mathbb{Z}^+ ; \\
\tilde f^k(t,s) = e^{-\frac{1}{2} \abs{k} s^2} H_n( \sqrt{\abs{k}} s ) g(t), \, g(t) \in L^2(\rr,\dif t) \text{ arbitrary.}
\end{align*}
$H_n$ are the Hermite polynomials, as seen in the wavefunction of the harmonic oscillator. Summing over the possible values of $k$, the spectrum of the left regular representation is $E \in [0,\infty)$.

\begin{table}
\centering
\begin{tabular}{| p{1.2cm} |*{4}{c|}} \hline
Bianchi & A.k.a & (a) $F(z)$ & (b) $\tilde{k}_0$, cross-section \\ \hline
II & Heisenberg & $\left( \begin{matrix} 1 & 0 \\ z & 1 \end{matrix} \right)$ & $\tilde k_0 = \left( \begin{matrix} 0 \\ k \end{matrix} \right),  k \in \mathbb{R}/\{0\} $\\
III & $\mathrm{Aff}(\mathbb{R}) \times \mathbb{R}$ & $\left( \begin{matrix} e^z & 0 \\ 0 & 1 \end{matrix} \right)$  & $\tilde k_0 = \left( \begin{matrix} k_2 \\ k \end{matrix} \right), (k,k_2) \in \mathbb{R} \times \{-1,+1\}$\\
IV &  & $\left( \begin{matrix} e^z & 0 \\ z e^z & e^z \end{matrix} \right)$ &$\tilde k_0 = \left( \begin{matrix} k \\ k_2 \end{matrix} \right), (k,k_2) \in \mathbb{R}_{+0} \times \{-1,+1\}$\\
V & & $\left( \begin{matrix} e^z & 0 \\ 0 & e^z \end{matrix} \right)$  & $\tilde k_0 = \left( \begin{matrix} \cos k \\ \sin k \end{matrix} \right), k \in \mathbb{R}/2\pi \mathbb{Z}$\\
VI \scriptsize{$q \in (-1,0)$} & & \multirow{2}{*}{$\left( \begin{matrix} e^z & 0 \\ 0 & e^{-q z} \end{matrix} \right)$} & $\tilde k_0 = \left( \begin{matrix} \cos k \\ \sin k \end{matrix} \right), k \in \mathbb{R}/2\pi \mathbb{Z}$\\
VI \scriptsize{$q \in (0,1)$}&  {\em e.g.} Poincar\'e & & $\tilde k_0 = \left( \begin{matrix} 0 & -1 \\ 1 & 0 \end{matrix} \right)^{k_2} \left( \begin{matrix} 1 \\ k \end{matrix} \right), (k,k_2) \in \mathbb{R}_{+0} \times \{0,1,2,3\}$\\
VII \scriptsize{$p \in \rr$} & {\em e.g.} Euclidean & $e^{pz} \left( \begin{matrix} \cos z & -\sin z \\ \sin z & \cos z \end{matrix} \right)$ &$\tilde k_0 = \left( \begin{matrix} k \\ 0 \end{matrix} \right), k \in (-e^{\pi p},-1] \cup [1,e^{\pi p}) $ \\ \hline
\end{tabular}

\begin{tabular}{|*{5}{c|}} \hline
Bi. & (c) Plancherel & (d) Automorphs & (e) Residual automorphs &
(f) $\mathcal{H}$ \\ \hline
II & $|k|$ & {\scriptsize $\left( \begin{matrix} a & 0 & c \\ d & aj-cg & f \\ g & 0 & j \end{matrix} \right)$}
& {\scriptsize $\lbrace \left( \begin{matrix} \cos \theta & 0 & \sin \theta \\ 0 & 1 & 0 \\ -\sin \theta & 0 & \cos \theta \end{matrix} \right),\left( \begin{matrix} 1 & 0 & 0 \\ 0 & -1 & 0 \\ 0 & 0 & -1 \end{matrix} \right) \rbrace$} & $X^2 + A Y^2 + Z^2$ \\
III &$1$& {\scriptsize $\left( \begin{matrix} a & 0 & c \\ 0 & e & f \\ 0 & 0 & 1 \end{matrix} \right)$}
 &{\scriptsize $\left( \begin{matrix} -1 & 0 & 0 \\ 0 & -1 & 0 \\ 0 & 0 & 1 \end{matrix} \right)$ }& $X^2 +  Y^2 + \alpha \{X,Y\} + A Z^2$ \\
IV &$1+k$& {\scriptsize $\left( \begin{matrix} a & 0 & c \\ d & a & f \\ 0 & 0 & 1 \end{matrix} \right)$}
 &{\scriptsize $\left( \begin{matrix} -1 & 0 & 0 \\ 0 & -1 & 0 \\ 0 & 0 & 1 \end{matrix} \right)$ }& $A(X^2 + Y^2) + B Z^2$ \\
V &$1$&{\scriptsize $\left( \begin{matrix} a & b & c \\ d & e & f \\ 0 & 0 & 1 \end{matrix} \right)$}
 & {\scriptsize $\lbrace \left( \begin{matrix} \cos \theta & \sin \theta & 0 \\ -\sin \theta & \cos \theta & 0 \\ 0 & 0 & 1 \end{matrix} \right),\left( \begin{matrix} 1 & 0 & 0 \\ 0 & -1 & 0 \\ 0 & 0 & 1 \end{matrix} \right) \rbrace$ }& $X^2 + Y^2 + A Z^2$ \\
VI& $\begin{cases} \cos^2 k - q \sin^2 k \\ q^{k_2 \mod 2} \end{cases}$& {\scriptsize $\left( \begin{matrix} a & 0 & c \\ 0 & e & f \\ 0 & 0 & 1 \end{matrix} \right)$}
 &{\scriptsize $\left( \begin{matrix} -1 & 0 & 0 \\ 0 & -1 & 0 \\ 0 & 0 & 1 \end{matrix} \right)$} & $X^2 +  Y^2 + \alpha \{X,Y\} + A Z^2$ \\
VII& $|k|$& {\scriptsize $\left( \begin{matrix} a & b & c \\ -b & a & f \\ 0 & 0 & 1 \end{matrix} \right)$}
 &{\scriptsize $\left( \begin{matrix} -1 & 0 & 0 \\ 0 & -1 & 0 \\ 0 & 0 & 1 \end{matrix} \right)$ }& $X^2 + A Y^2 + B Z^2$ \\ \hline
\end{tabular}
\captionsetup{singlelinecheck=off,font=small}
\caption[group structure]{The properties of the Bianchi groups II--VII. Results (a)--(d) from \cite{Avetisyan:2012zn}.\begin{enumerate}[(a)]
 \item $F(z)$ defines the group composition $(x,y,z) \circ (x^\prime,y^\prime,z^\prime) = (A_x,A_y,z+z^\prime)$, where $\left( \begin{matrix} A_x \\ A_y \end{matrix} \right) = \left( \begin{matrix} x \\ y \end{matrix} \right) + F(z) \left( \begin{matrix} x^\prime \\ y^\prime \end{matrix} \right)$.
 \item The irreps of the group, labelled by $k$ and sometimes also the discrete $k_2$. $D^{k(,k_2)}(s,t; \begin{smallmatrix} x \\ y \\ z \end{smallmatrix} ) = \exp(i \tilde k_0^T F(s) \tilde x) \del(s + z - t)$, where $\tilde x = (\begin{smallmatrix} x \\ y \end{smallmatrix})$.
 \item The Plancherel measure on the irreps. From the $k$ dependent part of $\abs{ \frac{ \pt F^T(s) \tilde k}{\pt s \pt k} }$.
 \item The automorphisms $M_{ij}$ of the Lie algebra, basis $X_j = (X,Y,Z)^T$, where $[M \xi, M \eta] = M[\xi,\eta], \, \forall \xi,\eta \in \mathfrak{g}$. All real values of the parameters are permitted, provided the matrix is non-singular. See also \cite{Harvey}.
 \item The subgroup of automorphs that leave the simplified Hamiltonian invariant.
 \item The simplified Hamiltonian.
\end{enumerate}
}
\label{tab:groupstructure}
\end{table}

\begin{table}
\centering
\begin{tabular}{| p{1.2cm} |*{2}{c|}} \hline
Bianchi & (a) ODE & DLMF \\ \hline
II & $f^{\prime \prime} + (E- A (k^2 + s^2)) f = 0 $ & \href{http://dlmf.nist.gov/12.2}{12.2.4}  \\
III & $A f^{\prime \prime} + (E - e^{2s} - k^2 - 2 \alpha k_2 k e^{s}) f = 0$ & \href{http://dlmf.nist.gov/13.14}{13.14} \\
IV & $B f^{\prime \prime} + (E - A e^{2s} (1+k^2 + 2 k_2 k s + k^2 s^2) ) f = 0 $ & - \\
V & $A f^{\prime \prime} + (E - e^{2s}) f = 0 $ & \href{http://dlmf.nist.gov/10.45}{10.45}  \\
VI \scriptsize{$q \in (-1,0)$} & $A f^{\prime \prime} + (E - \cos^2 k e^{2s} - \sin^2 k e^{-2 q s} - 2 \alpha \cos k \sin k e^{(1-q)s}) f = 0 $ & -\\
VI \scriptsize{$q \in (0,1)$} & $A f^{\prime \prime} + (E - e^{2s} - k^2 e^{-2 q s} - 2 \alpha k e^{(1-q)s}) f = 0 $ & -\\
VII \scriptsize{$p \in \rr$} & $B f^{\prime \prime} + (E - k^2 e^{2ps} (\cos^2 s + A \sin^2 s) ) f = 0$ & -\\ \hline
\end{tabular}
\begin{tabular}{| p{1.2cm} | p{5cm} | p{8cm} | } \hline
Bianchi & (b) Eigenvalues & (c) Solutions\\ \hline
II & $E = (2 n+1) \abs{k} + k^2 A$, $n \in \mathbb{Z}^+$ & $e^{-\frac{1}{2} \abs{k} s^2} H_n( \sqrt{\abs{k}} s )$ \\
III & $E \geq k_1^2$, and, $(n+\frac{1}{2}) \sqrt{A} + \sqrt{k^2 - E} = -\alpha k_2 k$, $n \in \mathbb{Z}^+$  & $e^{\frac{1}{2} s} W_{-\alpha k_1 k_2 A^{-\frac{1}{2}}, A^{-\frac{1}{2}} i \sqrt{E-k_1^2}} (2 A^{-\frac{1}{2}} e^{s})$, and $e^{\frac{1}{2} s} W_{-\alpha k_1 k_2 A^{-\frac{1}{2}}, A^{-\frac{1}{2}} \sqrt{k_1^2-E}} (2 A^{-\frac{1}{2}} e^{s})$, $e^{\frac{1}{2} s} M_{-\alpha k_1 k_2 A^{-\frac{1}{2}}, A^{-\frac{1}{2}} \sqrt{k_1^2-E}} (2 A^{-\frac{1}{2}} e^{s})$ \\
IV & $E \geq 0$ &- \\
V & $E \geq 0$ & $\tilde K_\frac{E}{A}(A^{-\frac{1}{2}} e^{s})$ \\
VI \scriptsize{$q \in (-1,0)$} &$E \geq 0$ &-\\
VI \scriptsize{$q \in (0,1)$} & $E$ discrete &-\\
VII \scriptsize{$p \in \rr$} & $E \geq 0$ &-\\ \hline
\end{tabular}
\captionsetup{singlelinecheck=off,font=normalsize}
\caption[spectrum]{\begin{enumerate}[(a)]
\item The ODEs coming from the action of $[\mathcal{H} + E]$ on the irreps. $f^\prime \equiv \frac{\dif}{\dif s} f$ etc.
\item The values of $E$ for the irrep $k(,k_2)$.
\item The analytic solutions $f(s)$ of the ODE (where possible) in the notation of the corresponding DLMF reference. $\tilde f^{k(,k_2)}(s,t) = f(s) g(t)$, where $g(t) \in L^2(\rr,\dif t)$ is arbitrary.
\end{enumerate}}
\label{tab:spectrum}
\end{table}

The spectra of the left regular representations of the other Bianchi
groups are obtained in same way. When decomposed into their actions on
the irreps, the Laplacians give second order ODEs in the form of a
Schr\"odinger equation, $f^{\prime \prime} + (\epsilon-V(s))f = 0$ for
some positive potential $V(s)$, and where $\epsilon \propto E$. Only
the ODEs of Bianchi II, III and V are analytically solvable, but we can still find the spectrum of each Bianchi group as follows.\footnote{In all cases, insisting that the Hamiltonian
  commute with one of the right-invariant vector fields results in an
  exactly-solvable model.} The potentials of Bianchi IV, VI ($q < 0$) and VII tend to zero as one of either $s \to \infty$ or $s \to - \infty$; the scattering theory of such one sided potentials predicts a continuum of solutions for $\epsilon > 0$ \cite{bender1999advanced}, and their degeneracies are given by the respective Plancherel measure. For Bianchi VI ($q>0$), $V(s) \to \infty$ as $s \to \pm \infty$, and we expect a infinite discrete set of `bound' solutions, the energies of which may be approximated by the WKB method, and are seen to tend smoothly to zero as $k \to 0$.\footnote{The nature of the WKB method as an asymptotic expansion of the exact energy eigenvalues means that, as the first term in the expansion for Bianchi VI ($q>0$) is an elliptic integral, the true values {\em cannot} be expressed in terms of simple functions.} Thus for all Bianchi groups, the set of all eigenvalues in the left regular representation is a continuum $E \geq 0$ and we know the density of these states in all but one case (Bianchi VI, $q>0$).

\subsection{Bianchi VIII --- $SL(2,\mathbb{R})$ \label{sec:b9}}
The Bianchi VIII algebra is semi-simple and isomorphic to the Lie
algebra of the group $SL(2,\rr)$. The latter group has fundamental
group $\pi_1 (SL(2,\rr)) = \zz$ and its universal cover is peculiar in
that it is not a matrix Lie group, having no finite-dimensional
faithful representations. The unirreps are given and the Plancherel
theorem derived in \cite{PUKANSZKY1964} and we simply quote the
results here, in the notation of \cite{bargmann1947}. The irreps are
given by the principal series $C^{(\tau)}_q$, with $0 \leq \tau <1$
and $q>\frac{1}{4}$, the two discrete series $D^\pm_l$, with
$l>0$, and those in the exceptional domain, $E^{(\tau)}_q$
with $0 \leq \tau <1$ and $\tau(1-\tau) < q \leq
\frac{1}{4}$. Denoting $\sigma = \sqrt{q-\frac{1}{4}}>0$ the
Plancherel decomposition is
\begin{multline}
f(g) = \int_0^\infty \dif \sigma \int_0^1 \dif \tau \sigma \mathrm{Re}
\tanh [\pi (\sigma + i\tau)] \mathrm{tr} [(C^{(\tau)}_q(g))^* \tilde{f}(\sigma, \tau)] +\\
+\sum_{\pm}\int_{\frac{1}{2}}^\infty \dif l (l - \frac{1}{2}) \mathrm{tr}
[(D^\pm_l(g))^*  \tilde{f}^\pm(l)].
\end{multline}
Note that only the $l$ irreps with $l>\frac{1}{2}$ appear.

Unfortunately the spectrum is not obtainable analytically, in general,
as we can see by focussing our attention on the irreps with $\tau \in
\{0,\frac{1}{2}\}$ (which are, in fact, the principal series irreps
appearing in the Plancherel measure on $SL(2,\rr)$). The Lie algebra ($[X-Y,H]=-2(X+Y)$, $[H,X+Y]=2(X-Y)$ and $[X+Y,X-Y] =-2H$, written in the $SU(1,1)$ basis) admits $SO(2,1)$ automorphisms, where the $SO(2)$ subgroup is $\{H,X+Y\}$. This simplifies the Hamiltonian to $\mathcal{H}=A(X-Y)^2 + B(X+Y)^2 + C H^2$, which on the principal irreps have action
\begin{equation*}
(A+e+g \cos(4 \phi) ) f^{\prime \prime} - g (i \sigma +2 ) \sin (4 \phi) f^\prime + (1+\sigma^2) (g \cos(4 \phi) -e) f = -E f
\end{equation*}
where $e=\frac{1}{2}(B+C)$, $g=\frac{1}{2}(B-C)$ and $f(\phi) \in
L^2(S^1, \dif \phi)$ and is odd in $\phi$ if $\tau=\frac{1}{2}$, and
even otherwise. $\phi \in (-\pi,\pi]$. When the ODE is put in normal form, the
first order term in the WKB approximation to $E$ reduces to an
elliptic integral, implying that the (`all-order') exact result is not
a simple function of exponentials, logarithms and powers. However, in
the symmetric limit $g\to 0$ ({\em i.e.} $B\to C$), the ODE and
spectrum for these irreps tend to those of a free particle on $S^1$.
\subsection{Bianchi IX --- $SU(2)$  \label{sec:b8}}

\newcommand{\be}{\beta}
\newcommand{\ga}{\gamma}
\newcommand{\pta}{\pt_\alpha}
\newcommand{\ptb}{\pt_\beta}
\newcommand{\ptc}{\pt_\gamma}

Finally, we turn to the archetypical rigid body, starting our
  discussion with the more familiar case of $SO(3)$, before addressing
  its universal cover, $SU(2)$. Examples abound in nature of systems
whose configuration is described by an arbitrary rotation, and whose
Hamiltonian is also rotation invariant. Indeed, via spinning tops and
the rotational spectroscopy of gases, both the classical and quantum
dynamics of the rigid body have been well studied. We calculate the spectrum of the quantum mechanical rigid body using the formalism of the previous Sections, and show it to be consistent with known results.

The $SO(3)$ group coordinates are typically given in terms of Euler
angles $\al,\be,\ga$ (we use here the z-y-z `active'
convention \cite{biedenharn2009angular}), where $\al,\ga \in [0,2 \pi]$, $\be \in
[0,\pi]$. The Haar measure is $\frac{\dif \al \dif (\cos \be) \dif
  \ga}{8 \pi^2}$. The unirrep matrices are the Wigner D-matrices
$D^{j}_{mk}(\begin{smallmatrix} \al \\ \be \\ \ga \end{smallmatrix})$,
labelled by $j\in \{0,1,2\dots \}$, where $m,k \in \{-j,-j+1,\dots,j\}$ are the indices of the resulting $(2j+1)\times(2j+1)$ matrix. The explicit form of $D^{j}_{mk}$, and the composition of two sets of Euler angles are, in general, complicated. We refer the reader to \cite{biedenharn2009angular}, from which we quote the following results.

The Wigner D-matrices form, respectively, a Schur orthogonality relation on the irreps, and a completeness relation on the group coordinates:
\begin{align*}
\int \frac{\dif \al \dif \cos \be \dif \ga}{8 \pi^2} \overline{D}^j_{km}(\begin{smallmatrix} \al \\ \be \\ \ga \end{smallmatrix}) D^{j^\prime}_{m^\prime,k^\prime}(\begin{smallmatrix} \al \\ \be \\ \ga \end{smallmatrix}) &= \frac{1}{2j+1} \delta^{j j^\prime} \delta^{m m^\prime} \delta^{k k^\prime} ,\\
\sum_{j,m,k}  (2j+1) D^{j}_{mk}(\begin{smallmatrix} \al \\ \be \\ \ga \end{smallmatrix}) \overline{D}^j_{km}(\begin{smallmatrix} \al^\prime \\ \be^\prime \\ \ga^\prime \end{smallmatrix}) &= 8 \pi^2 \delta(\al-\al^\prime) \delta(\cos \be - \cos \be^\prime) \delta(\ga - \ga^\prime) ,
\end{align*}
where $\sum_{j,m,k} = \sum_{j=0}^\infty \sum_{m=-j}^{+j}
\sum_{k=-j}^{+j}$. Using the same notation, one may decompose the left
regular representation via the decomposition of functions $f \in L^2(SO(3))$:
\begin{align*}
 f(\begin{smallmatrix} \al \\ \be \\ \ga \end{smallmatrix}) &= \sum_{j,m,k}  (2j+1) D^{j}_{mk}(\begin{smallmatrix} \al \\ \be \\ \ga \end{smallmatrix}) \tilde f^j_{km} , \\
 \tilde f^j_{mk} &= \int \frac{\dif \al \dif \cos \be \dif \ga}{8 \pi^2} \overline{D}^{j}_{mk}(\begin{smallmatrix} \al \\ \be \\ \ga \end{smallmatrix}) f(\begin{smallmatrix} \al \\ \be \\ \ga \end{smallmatrix}).
\end{align*}
Note that, for a given value of $j$, there are $(2j+1)$ vectors
$\tilde f^j_{mk}$ (labelled by $k \in \{-j,-j+1,\dots,j\}$) in the
decomposition of $f$ which, under the group action, transform
under the $j$th irrep of $SO(3)$. This is a manifestation of a general
result for compact groups, the Peter-Weyl theorem
\cite{peterweyl,BarutR}, where an irrep of dimension $d$ features $d$
times in the Hilbert space of the left regular representation. The
  Plancherel measure is thus $2j+1$ times the counting measure on $j
  \in \{0,1,2,\dots\}$. These
$2j+1$ copies are degenerate in energy in the rotationally-invariant
systems we consider, but are easily observed, {\em e.g.} in the rotation spectra of
molecules in the presence of an electric field \cite{GordyCook}.

Having obtained the decomposition of the regular representation, we
again proceed to construct the Hamiltonian. The structure constants of the Lie algebra are
$\epsilon_{ijk}$ and the automorphism group is immediately seen to be
$SO(3)$. We may thus pick the special orthogonal transformation
necessary to diagonalize the real symmetric matrix of the Hamiltonian coefficients, leaving $\mathcal{H} = A J_1^2 + B J_2^2 + C J_3^2$
for some (positive) eigenvalues $A$, $B$ and $C$.\footnote{In other
words, we can pick principal axes with respect to which the moment
of inertia tensor is diagonal.} The residual automorphisms are the group of order 4 generated by $\{e^{\pi J_i}\}$, isomorphic to the Klein
group, whose effect is to change the sign of two
of the three $J_i$. The Klein group is Abelian, with 4, 1-d
irreps. The permuting automorphisms are the group of order 24 generated by $\{e^{\frac{\pi}{2} J_i}\}$, isomorphic to the symmetry group of a
cube, with 2 1-d, 1 2-d, and 2 3-d irreps.

The action of $\mathcal{H}$ on the left regular representation is via
the self-adjoint right-invariant vectors
\begin{align*}
J^R_1 = ( \cot \be \cos \al \pta + \sin \al \ptb - \frac{\cos \al}{\sin \be} \ptc  ) ,\\
J^R_2 = ( \cot \be \sin \al \pta - \cos \al \ptb - \frac{\sin \al}{\sin \be} \ptc  ),\\
J^R_3 = -  \pta .
\end{align*}
Making use of the properties
\begin{align*}
J^R_3 D^j_{mk}(\begin{smallmatrix} \al \\ \be \\ \ga \end{smallmatrix}) = i m D^j_{mk}(\begin{smallmatrix} \al \\ \be \\ \ga \end{smallmatrix}) \text{ and } \\
(J^R_1 \pm i J^R_2) D^j_{mk}(\begin{smallmatrix} \al \\ \be \\ \ga \end{smallmatrix}) = i \sqrt{(j \pm m)(j \mp m+1)} D^j_{m \mp 1,k}(\begin{smallmatrix} \al \\ \be \\ \ga \end{smallmatrix}),
\end{align*}
we find the action of $\mathcal{H}$ on the irreps to be
\begin{equation*}
 \mathcal{H} f(\begin{smallmatrix} \al \\ \be \\ \ga \end{smallmatrix}) =  - \sum_{j,m,m^\prime,k}  (2j+1) D^{j}_{mk}(\begin{smallmatrix} \al \\ \be \\ \ga \end{smallmatrix}) [ A {J^j_1}^2 + B {J^j_2}^2 + C {J^j_3}^2 ]_{m m^\prime} \tilde f^j_{m^\prime k} ,
\end{equation*}
where $J^j_i$ is the canonical form of the angular momentum generator
in the $j$th matrix rep. Finding the spectrum of the left regular rep
then amounts to diagonalizing a real symmetric matrix, whose elements
are linear combinations of $A$, $B$ and $C$, for each $j$. This
diagonalization is simplified by first decomposing with respect to the
Klein group of residual automorphisms. Thus, the 3-d $j=1$ irrep
decomposes into one copy of each of the non-trivial irreps of the
Klein group and the spectrum is found to be
$E =
\frac{A+B}{2}$, $\frac{B+C}{2},$ or $\frac{C+A}{2}$, with a 3-fold
degeneracy of each.\footnote{The $j=1$ irrep `decomposes' into one of
  the 3-d irreps of the cube group of permuting automorphisms and so
  we find that we can obtain all of the energy eigenvalues from
  any one by permutations of $A,B$, and $C$.} For $j=2$, the 5-d irrep decomposes into one copy of each of
the non-trivial irreps of the Klein group and two copies of the
trivial irrep. By Schur's lemmata, the latter two can mix and finding
the spectrum requires us to diagonalize a $2 \times 2$ matrix. Unfortunately, no closed form is available for the spectrum
at arbitrary $j$. However, thus computed, these energies (and indeed the degeneracies,
measured by the strength of spectral features) are precisely those
observed in molecular microwave
spectra~\cite{GordyCook,kroto2003molecular}, where $A$, $B$ and $C$
are reciprocals of twice the principal moments of
inertia of the molecule.

The group $SO(3)$ furnishes us also with the archetype of a coset representation. This is seen, for example, in a linear molecule (such as CO or CO$_2$, but not H$_2$O) where the system's configurations under a rotation about a given axis are equivalent. The orientation of the linear molecule is then instead given by the two angles of $SO(3)/SO(2) \cong S^2$.

We choose $\al$ and $\be$ to be the two remaining coordinates, whereas $\ga$ parametrises the coset $SO(2)$. We can make a closed, $SO(3)$-invariant volume form on the group, namely $(\cos \ga \sin \be \dif \al - \sin \ga \dif \be) \wedge (-\sin \ga \sin \be \dif \al - \cos \ga \dif \be) = -\sin \be \dif \al \wedge \dif \be$, from two left-invariant one forms. Let the measure be $\frac{\sin \be \dif \al \dif \be}{4 \pi}$.

The matrices $D^j_{mk}(\begin{smallmatrix} \al \\ \be \\ 0 \end{smallmatrix})$ are proportional to the spin-weighted spherical harmonics on $S^2$, parametrised by spherical polar coordinates ($\be$,$\al$).\footnote{In the simplest case $k=0$, $D^l_{m0}(\begin{smallmatrix} \al \\ \be \\ 0 \end{smallmatrix}) = \sqrt{\frac{4 \pi}{2l+1}} \overline{Y}_{l,m}(\be,\al)$, the usual spherical harmonic functions.} For each $k$, they form a complete set of functions on $S^2$, leading to the relation
\begin{align*}
\sum_{j,m,k} D^{j}_{mk}(\begin{smallmatrix} \al \\ \be \\ 0 \end{smallmatrix}) \overline{D}^j_{km}(\begin{smallmatrix} \al^\prime \\ \be^\prime \\ 0 \end{smallmatrix}) &= 4 \pi \delta(\cos \be - \cos \be^\prime) \delta(\al-\al^\prime) .
\end{align*}
The coset representation decomposes thus:
\begin{align*}
 f(\begin{smallmatrix} \al \\ \be \end{smallmatrix}) &= \sum_{j,m,k}  D^{j}_{mk}(\begin{smallmatrix} \al \\ \be \\ 0 \end{smallmatrix}) \tilde f^j_{km} , \\
 \tilde f^j_{mk} &= \int \frac{\dif \cos \be \dif \al}{4 \pi} \overline{D}^{j}_{mk}(\begin{smallmatrix} \al \\ \be \\ 0 \end{smallmatrix}) f(\begin{smallmatrix} \al \\ \be \end{smallmatrix}).
\end{align*}

Note that the extra degeneracy of the higher dimension irreps is
gone. The coset representation contains each $j$ irrep exactly once and the
  Plancherel measure is now just the counting measure on $j
  \in \{0,1,2,\dots\}$.

The projection to the coset space is given by $\pi(\begin{smallmatrix} \al \\ \be \\
  \ga \end{smallmatrix}) = (\begin{smallmatrix} \al \\
  \be \end{smallmatrix})$, which gives $\pi^*(\pt_\al) = \pt_\al$,
$\pi^*(\pt_\be) = \pt_\be$ and $\pi^*(\pt_\ga) = 0$. The linear rigid rotor
 Hamiltonian is given by
\begin{gather}
\mathcal{H} = A((J^R_1)^2 + (J^R_2)^2+(J^R_3)^2),
\end{gather}
where $J^R_i$ are now the pushforwards to $S^2$ of the right-invariant
vector fields on $SO(3)$. We find
\begin{gather}
\mathcal{H} = A \left[ \pt_\be^2 + \cot \be \pt_\be + \frac{1}{\sin^2 \be} \pt_\al^2 \right].
\end{gather}
The spectrum is that of the Laplacian in spherical polar coordinates. This can equally be seen via the decomposition, which contains a piece
\begin{gather*}
 \mathcal{H} f(\begin{smallmatrix} \al \\ \be \end{smallmatrix}) \supset  \sum_{j,m}  -[A j(j+1) ] D^{j}_{m0}(\begin{smallmatrix} \al \\ \be \\ 0 \end{smallmatrix})  \tilde f^j_{0m},
\end{gather*}
which simply
multiplies each vector $\tilde f^j_{0m}$ by $A
j(j+1)$. The action on the $\tilde f^j_{km}, k \neq 0$ is more
complicated, but does not yield eigenfunctions. Thus the possible states have energy $A j(j+1)$ and degeneracy $2j+1$ for each $j
  \in \{0,1,2,\dots\}$, as seen in the heat capacity of molecular
  hydrogen.

The results for the case of $SU(2)$ are analogous and easily obtained. The
irreps of $SU(2)$ contain, in addition to those of $SO(3)$, the irreps
with $j \in \{\frac{1}{2},\frac{3}{2},\dots\}$. The Plancherel measure
is $2j+1$ times the counting measure on $j \in
\{0,\frac{1}{2},1,\frac{3}{2},\dots\}$, by Peter-Weyl. The groups of
residual and permuting automorphisms are the preimages under the
projection map $\pi: SU(2) \rightarrow SO(3)$ of the corresponding
groups for $SO(3)$. The residual group of automorphisms of $SU(2)$ is
isomorphic to the (non-Abelian) quaternion group of order 8 \cite{DowkerPett} and features, in addition to the 4
1-d irreps of the Klein group, a 2-d irrep.\footnote{The permuting
  group is of order 48 and has, in addition to the irreps of the cube
  group, 2 2-d irreps and 1 4-d irrep.} The $j=\frac{1}{2}$ irrep
of $SU(2)$ decomposes into the 2-d irrep. The energy eigenvalue
$\frac{A+B+C}{4}$ is 4-fold degenerate, with a 2-fold
degeneracy coming from rotational invariance via the Plancherel
measure and another 2-fold degeneracy that can be understood as a
Kramers degeneracy arising from time-reversal invariance.
\section{Discussion}
We have given a formulation of the quantum mechanics of a
particle on a Lie group manifold $G$, whose classical motions follow
geodesics of an arbitrary one-sided-invariant metric. We have also shown
how this formulation can sometimes be extended to motion on a coset
space $G/H$. In many cases, the system turns out to be exactly
solvable using standard methods of harmonic analysis and
representation theory; we have also described the various obstacles to
doing so. 

We have studied all Lie algebras up to dimension three as examples.
The prototypical example of a rigid body is somewhat atypical, since
it is the only case in $d \leq 3$ that is simply-connected and also
compact, with finite-dimensional unitary, irreducible representations.
In non-compact cases, the spectrum is continuous and (at least
in the cases 
where we are able to compute it) takes the simple form $E \geq 0$. This
does not mean, of course, that the physics of these different systems
is the same. Indeed, what is important for the physics is not just the
spectrum of eigenvalues, but also the degeneracy of those eigenvalues
--- the density of states, in physicists' language. The degeneracies,
given by the Plancherel measure in each case, are markedly different
and we expect that they will lead to markedly different physics. 

It would be of interest to try to reproduce our results using path
integral methods.\footnote{For the rigid body case, see \cite{schulman}.} We expect this to be non-trivial, even in the
case where $G$ is compact, where the quantum mechanics on $G/H$
can be described by the non-linear sigma model in 0+1-d with target
space $G/H$. Indeed, just as for the quantum mechanics of a
free particle on $\rr^n$ (where the generalised eigenfunctions
$e^{ik\cdot x}$
correspond to a uniform probability density for finding the particle
anywhere in $\rr^n$), the energy eigenstates for general $G/H$ are delocalised
(in the sense that the moduli-squared of the generalised
eigenfunctions do not tend to 0 at infinity) in all
cases where we have obtained their explicit form. 
This is in accord with our earlier observation that attempting to evaluate the path
integral by assuming that the particle is localised leads to a
contradiction. So, evaluation of the path integral will need to take
account of global considerations.

Whilst these exactly-solvable, free-particle models may seem somewhat esoteric, one
  should bear in mind that the notion of dynamics on a manifold with
  invariance under a
  Lie group action is a rather general one in physics. We anticipate that the exactly-solvable models provide a useful starting point for
doing quantum-mechanical perturbation theory for dynamical systems of
this type, generalising the use of the harmonic oscillator and free
particle. Indeed, the rigid body can be used as the starting point for
a perturbative study centrifugal distortions of polyatomic molecules
\cite{GordyCook}.

The quantum mechanical systems of a rigid body and a free particle, corresponding to $G = SO(3)$ and
$G=\rr^n$ are, of course, ubiquitous in physics and chemistry. Do
there exist dynamical systems to which we can apply the results for
other $G$? To give just one example, it has been suggested that the
dynamics of a certain ellipsoidal vortex solution in a fluid can,
under certain restrictions, be modelled as a
system moving on $SO(2,1)$ \cite{kida}. 

The most interesting application, from our point of view, is to the
quantisation of a perfect, incompressible fluid moving on a manifold $M$. This corresponds to $G
= \mathrm{SDiff} (M)$, which is infinite dimensional. Two ways to
proceed suggest themselves. The first is to consider an $M$ which
admits a complex structure, for which one can consider a simplified
toy model of a fluid, in which one insists that the fluid maps be not
just smooth, but rather be analytic. $\mathrm{SDiff} (M)$ is then replaced by the group of
(volume-preserving) conformal
holomorphisms of $M$, which is finite-dimensional and can be quantised
using the methods described here. These simplified
systems (which we propose to call `analytic fluids') may present a
useful intermediate step on the road to quantising a genuine
incompressible fluid.

The second approach is to try to define the infinite-dimensional group
$\mathrm{SDiff} (M)$, for some $M$, as a limit of a sequence of
finite-dimensional groups. For the case where $M$ is either a 2-sphere
or a 2-torus (which
bore so much fruit for Arnold at the classical level), it has been
observed \cite{goldstone,Hoppe:1986aj} that the structure constants of
the Lie algebra of $\mathrm{SDiff} (M)$ can be
obtained (in a specific basis) as a convergent sequence of those of
$SU(N)$, as $N \rightarrow \infty$. Unfortunately, it is known \cite{Bordemann:1990pa} that
these algebras are not isomorphic, so it is unclear whether this
observation can be used in a mathematically-sound way.
\section*{Acknowledgements}
We thank J.~Davighi, J.~Holgate, A.~Lamacraft, C.~Mouhot, M.~Nardecchia,
O.~Randal-Williams, and R.~Rattazzi for discussions. BG acknowledges
the support of STFC (grant ST/L000385/1) and King's College,
Cambridge. DS also acknowledges the support of STFC, and Emmanuel College, Cambridge.

\appendix
\section{Weyl sequences for the \affr~ group\label{app:weylseq}}

We wish to show that the spectrum of the differential operator $L = x^2 \frac{\dif^2}{\dif x^2}+x \frac{\dif}{\dif x}-x^2$, in the function space $L^2(\frac{\dif x}{x},\rr^+)$, is indeed $E \in [0,\infty)$, as stated in \S\ref{sec:leftspec}. To do so, for each possible eigenvalue $E \equiv \nu^2$ we construct a Weyl sequence out of the corresponding generalised eigenfunction $\tilde K_\nu(x) \equiv K_{i \nu}(x)$. A Weyl sequence is a set of functions $\{w_n : w_n \in L^2(\frac{\dif x}{x},\rr^+), \forall n \in \nn \}$, all of which have unit $L^2$ norm ($\abs{\abs{w_n}} = 1$), and for which $\abs{\abs{(L+\nu^2)w_n}} \to 0$ as $n \to \infty$. It will suffice to show that
\begin{equation}
w_n = A_n x^\frac{1}{n} \tilde K_\nu(x) , 
\end{equation}
(where $A_n$ is the appropriate prefactor to ensure $\abs{\abs{w_n}} = 1, \forall n \in \mathbb{N}$) satisfies these requirements for any $\nu \geq 0$, and hence $E \in [0,\infty)$ by exhaustion.

To compute the norm of the $w_n$, we need to evaluate the integrals
\begin{equation*}
F(\lambda,\nu) = \int_0^\infty \dif x \, x^{-\lambda} \abs{\tilde K_\nu(x)}^2 .
\end{equation*}
The properties of $K_\mu(x)$ are such that it is real when $\mu$ and
$x$ are both real and positive \cite[10.45]{DLMF}. Moreover, $K_\mu(x)$, when $x \neq 0$, is entire in $\mu$. We may therefore assert, using Schwartz's reflection principle, that $K_\mu(x)^* = K_{\mu^*} (x)$ for real $x$. Thus $\tilde K_\nu(x)^* = \tilde K_{-\nu}(x)$, and
\begin{equation}
F(\lambda,\nu) = \int_0^\infty \dif x \, x^{-\lambda} K_{i\nu}(x) K_{-i\nu}(x) = \frac{1}{4} \sqrt{\pi} \frac{ \Gamma(\frac{1}{2}-\frac{\lambda}{2}) }{ \Gamma(1-\frac{\lambda}{2})} \Gamma(\frac{1}{2}-\frac{\lambda}{2}+ i \nu) \Gamma(\frac{1}{2}-\frac{\lambda}{2} - i \nu) \text{ where $\lambda<1$},
\end{equation}
from \cite[6.576.4]{gradshteyn2014table}. The choice $A_n^2 = F(1-\frac{2}{n},\nu)^{-1}$ guarantees $\abs{\abs{w_n}}=1$. To evaluate $\abs{\abs{(L+\nu^2)w_n}}$, first use integration by parts and the Bessel function's ODE to write the following integrals in terms of $F$:
\begin{align*}
\int_0^\infty \dif x \, x^{-\lambda} [ K^\prime_{i \nu}(x) K_{-i \nu}(x) + K_{i \nu}(x) K^\prime_{-i \nu}(x) ] = \lambda F(\lambda+1,\nu) \text{   where $\lambda<0$},\\
\int_0^\infty \dif x \, x^{-\lambda}  K^\prime_{i \nu}(x) K^\prime_{-i \nu}(x) = [\nu^2 + \frac{1}{2} (\lambda+1)^2 ] F(\lambda+2,\nu) - F(\lambda,\nu) \text{   where $\lambda<-1$} .
\end{align*}
Then, by writing
\begin{equation}
A_n^{-1} (L+\nu^2) w_n = \frac{2}{n} x^{1+\frac{1}{n}} K^\prime_{i \nu} (x) + \frac{1}{n^2} x^{\frac{1}{n}} K_{i \nu} (x) ,
\end{equation}
we obtain
\begin{align}
\abs{\abs{(L+\nu^2)w_n}} &= A_n^2 \frac{4}{n^2} \int_0^\infty \dif x \, x^{1+\frac{2}{n}}  K^\prime_{i \nu}(x) K^\prime_{-i \nu}(x) \nonumber \\
&+ A_n^2 \frac{2}{n^3} \int_0^\infty \dif x \, x^{\frac{2}{n}} [ K^\prime_{i \nu}(x) K_{-i \nu}(x) + K_{i \nu}(x) K^\prime_{-i \nu}(x) ]  \nonumber \\
&+ A_n^2 \frac{1}{n^4} \int_0^\infty \dif x \, x^{-1 + \frac{2}{n}} K_{i \nu}(x) K_{-i \nu}(x)  \nonumber \\
&= \frac{4 \nu^2}{n^2} + \frac{5}{n^4} - \frac{4}{n^2} \frac{F(-1-\frac{2}{n},\nu)}{F(1-\frac{2}{n},\nu)} .
\end{align}
The ratio of gamma functions implicit in the last line simplifies to
\begin{equation}
\abs{\abs{(L+\nu^2)w_n}} = \frac{4 \nu^2 n^3 + 5 n + 2}{(2+n) n^4} \stackrel{n \to \infty}{\rightarrow} 0 .
\end{equation}

We have therefore found a Weyl sequence for every eigenvalue $E \in [0,\infty)$.

\bibliography{rigid.bib}
\bibliographystyle{JHEP}

\end{document}